\documentclass[12pt]{article}

\usepackage{amsmath,amsthm, amsfonts, amssymb, amsxtra, amsopn}
\usepackage{pgfplots}
\usepgfplotslibrary{colorbrewer}
\pgfplotsset{compat = 1.15, 
			 cycle list/Set1-3} 
\usetikzlibrary{pgfplots.statistics, pgfplots.colorbrewer} 
\usepackage{pgfplotstable}
\usepackage{graphicx,grffile}
\usepackage{multirow}
\usepackage{booktabs}
\usepackage{tcolorbox}
\usepackage{algorithm} % Used for writing algorithms in a paper
\usepackage[noend]{algpseudocode} % Allows psuedocode keywords (e.g., "if", "while", "for", etc.) in algorithms.
\usepackage{listings}
\usepackage{cmap}
\usepackage{colortbl}
\usepackage{adjustbox}
\usepackage{epsfig}

\usepackage[tableposition=top,font=small,skip=5pt]{caption}
\usepackage{subcaption}
\usepackage{makecell}
\usepackage[explicit]{titlesec}

\usetikzlibrary{patterns}

\def\bbb#1{{\color{blue}#1}}
\def\com#1{\bbb{\texttt{/\kern-1.5pt /} #1}}

\def\tau{\mathcal{T}}

\PassOptionsToPackage{hyphens}{url}
\PassOptionsToPackage{table}{xcolor}

\usepackage{hyperref}
\hypersetup{colorlinks=true,linkcolor=black,citecolor=black,urlcolor=blue,filecolor=black}
\hypersetup{pdfpagemode=UseNone,pdfstartview=}

\definecolor{darkgreen}{rgb}{0.125,0.5,0.169}
\usepackage[shortlabels]{enumitem}
\setlist[itemize]{noitemsep, topsep=0pt}

\advance\oddsidemargin by -0.45in
\advance\textwidth by 0.9in

\advance\topmargin by -0.6in
\advance\textheight by 1.2in

\long\def\symbolfootnotetext[#1]#2{\begingroup%
  \def\thefootnote{\fnsymbol{footnote}}\footnotetext[#1]{#2}\endgroup}

\newcommand\dunderline[3][-1pt]{{%
      \sbox0{#3}%
      \ooalign{\copy0\cr\rule[\dimexpr#1-#2\relax]{\wd0}{#2}}}}
\def\uuu{\kern-1pt\dunderline{0.75pt}{\phantom{M}}}

\def\grayscale{\textrm{Grayscale}}
\def\entropyHilbert{\textrm{Entropy Hilbert}}
\def\byteclassHilbert{\textrm{Byteclass Hilbert}}
\def\hit{\textrm{HIT}}
\def\bigramCartesian{\textrm{Bigram Cartesian}}
\def\bigramPolar{\textrm{Bigram Polar}}
\def\spiral{\textrm{Spiral}}
\def\byteclass{\textrm{Byteclass}}

\DeclareMathOperator{\thth}{th}

\def\zz{\phantom{0}}

\def\un{\raisebox{1pt}{\underline{\phantom{n}}}}

\title{A Comparison of Malware Image Transformations Using Grad-CAM and Hybrid Learning Models}

\author{Vibha Bhavikatti\footnotemark[1]\ \ \ 
Mark Stamp\footnotemark[1]\,\,\footnotemark[2]} 

\begin{document}

\symbolfootnotetext[1]{Department of Computer Science, San Jose State University}
\symbolfootnotetext[2]{mark.stamp$@$sjsu.edu}

\maketitle

\abstract
Recent studies have shown that binary-to-image representations can enable 
effective machine learning-based results for malware detection and classification. 
However, performance
can vary significantly, depending on the technique used to convert binaries to images. 
Furthermore, the explainability and interpretability of image-based models 
is largely unexplored within the malware domain. 
In this research, we employ Gradient-weighted Class Activation Maps (Grad-CAM)
as an eXplainable AI (XAI) tool, which we use to analyze
eight distinct image types derived from malware samples.
We provide quantitative faithfulness and stability 
metrics for Grad-CAM heatmaps and we compare these heatmaps 
to High-Resolution Class Activation Mappings (HiResCAM). 
We also show that Grad-CAM heatmaps can provide useful information
for malware classification. 
Specifically, we show that a Random Forest model 
trained on features extracted from Grad-CAM images 
via a MobileNetV2  Convolutional Neural Network (CNN) model achieves a test accuracy 
of~0.777 across~17 malware families, exceeding a previous benchmark of~0.750 
for this same dataset. A key finding of this research is that for the malware
image transformations considered, accuracy and explanation faithfulness do not coincide, 
e.g., image transformation techniques that produce the most faithful explanations yield 
only mid-tier accuracy.

\bigskip


\section{Introduction}

Malware continues to be one of the most significant threats to modern computing systems, 
with malware families evolving to evade traditional signature-based defenses. 
As a result, machine learning-based malware classification has become a major research 
direction, enabling automated analysis at scale using behavioral, static, or hybrid features 
extracted from binaries~\cite{DamodaranS17}. However, many high-performing 
models---especially Deep Neural Networks (DNN)---are difficult for analysts to interpret, which 
limits trust and may therefore limit adoption in security applications.

An approach that has proven highly successful is to convert malware binaries into images, 
so that patterns can be detected using advanced computer vision techniques. Early image-based
approaches mapped
raw bytes or opcodes directly into grayscale images, then applied 
Convolutional Neural Networks (CNN) or crafted image descriptors 
for family classification~\cite{nataraj2011malware}.

Subsequent research has explored more diverse encodings, such as space-filling curves 
(e.g., Hilbert or spiral curves), bigram frequency maps, entropy-based visualizations, 
and hybrid spatial-frequency representations to highlight structural and statistical characteristics 
of malware binaries~\cite{stamp2024malwareimages,entropy2022}. 
These transformations often produce distinct textures 
for different families, making them appealing for both improved accuracy and visual inspection.

At the same time, there is growing interest in eXplainable AI (XAI) in malware analysis, 
so that security analysts can understand why a model predicts a given family or label. 
Methods such as Gradient-weighted Class Activation Maps (Grad-CAM) have been widely used in 
computer vision to produce heatmaps over input images, which serve to indicate which 
regions drive a model’s decision~\cite{selvaraju2017gradcam}.

Recent work has applied Grad-CAM and related XAI techniques to malware images, 
showing that explanations can reveal whether the model focuses on meaningful structural 
regions or spurious artifacts~\cite{xai}. However, there is still limited understanding of how 
different types of malware image transformations affect both classification performance and the 
structure of Grad-CAM explanations and similar XAI techniques.

The research presented in this chapter builds on prior work by 
Agrawal et al.~\cite{stamp2024malwareimages}, where 
eight distinct exe-to-image transformation techniques
are considered. We aim to systematically study how 
these different transformation techniques influence both classification performance and 
model explainability. Using Grad-CAM heatmaps, we analyze where machine learning models 
focus during prediction, and investigate whether these attention maps contain consistent, 
class-specific information. This chapter additionally explores whether Grad-CAM patterns 
themselves can be used for downstream tasks, such as malware family classification.
%or transformation classification.

In the prior study~\cite{stamp2024malwareimages}, each of eight distinct image transformation 
techniques was evaluated using Histogram of Oriented Gradients (HOG) features, 
color and texture descriptors, statistical features, 
and classical machine learning classifiers including 
Support Vector Machines (SVM), Random Forest, and XGBoost. The key 
finding was that different transformations preserve different levels of discriminative family structure, 
with the best accuracy achieved on a~17-family malware dataset being approximately~75\%.

In this chapter, we build on the previous work in~\cite{stamp2024malwareimages} 
to investigate how CNNs behave across different malware-to-image 
transformations, how Grad-CAM explanation heatmaps differ between transformation types, 
how explanations vary per malware family, and how hybrid feature 
combinations derived from CNN embeddings perform as a feature engineering step.
Specifically, the research presented in this chapter consists of the following.
\begin{itemize}
    \item We generate Grad-CAM heatmaps for each malware sample over eight 
    malware-to-image transformations, and we consider how the 
    visual patterns differ across families and transformations.
%    \item Large-scale Grad-CAM dataset generation across multiple image transformations 
%    and malware families.
    \item We treat these Grad-CAM heatmaps as additional malware images and perform feature extraction 
    %(statistical moments, region geometry, LBP, GLCM, Hu moments) 
    to assess whether these XAI images 
    themselves preserve enough structure to support family classification. % using classical models.
    %(Random Forest, XGBoost, etc.).
    \item We compare CNN-based and feature-based approaches on the generated Grad-CAM 
    dataset and analyze which transformations yield the best trade-off between 
    accuracy and interpretability.
    \item We generate HiResCAM heatmaps to evaluate whether this higher-resolution 
    XAI method produces meaningfully better results than Grad-CAM on malware images.
    \item We quantitatively evaluate Grad-CAM explanations using faithfulness 
    and stability metrics across all eight transformation types.
    \item We compare classification performance between models trained on the original images 
    and models trained on Grad-CAM overlay images.
    \item We train various models on feature combinations derived from MobileNetV2 embeddings, thereby
    improving on a benchmark result for malware family classification.
\end{itemize}

In summary, by systematically evaluating both performance and explainability across 
multiple malware image transformations and Grad-CAM pipelines, we aim to clarify the 
role of visualization choices in interpretable malware classification. We also explore whether 
explanations themselves can be used as a secondary representation to improve model accuracy.

The remainder of this chapter is organized as follows. Section~\ref{chap:background} 
provides relevant background information, including details on the dataset.
Section~\ref{chap:methodology} covers various implementation details. 
In Section~\ref{chap:results}, we present our experimental results, including 
classification performance, along with qualitative and quantitative analysis of Grad-CAM 
explanations. Section~\ref{chap:discussion} discusses the implications of these results, 
including the trade-offs between accuracy and interpretability. 
Finally, Section~\ref{chap:conclusion} summarizes the contributions of this work and 
outlines directions for future research.

\section{Background}\label{chap:background}

Traditional malware analysis relies on fixed code patterns or observed runtime behavior.
These methods have trouble when malware is obfuscated, packed, or code is polymorphic. 
To address these limitations, researchers have successfully used machine learning models for at least
the past~20 years~\cite{wing}, and more recently, deep learning techniques having shown considerable 
success~\cite{Stamp2021}. In recent years, image-based malware analysis---where malware samples 
are converted into 2-D images---has become a dominant
trend in malware research~\cite{Bhodia19,nataraj2011malware,Yajamanam18}. 
This has enabled the use of advanced computer vision techniques, 
which often perform surprisingly well in the malware context.

Early work by Nataraj et al.~\cite{nataraj2011malware} showed that mapping raw bytes directly to 
grayscale images produces distinctive visual textures that correlate strongly with malware families. 
Since then, many binary-to-image transformations have been proposed, each capturing different 
statistical or spatial properties of a malware sample. 
These transformations vary widely in how they 
preserve structure, entropy, locality of bytes, and code layout~\cite{stamp2024malwareimages}.

\subsection{Malware Dataset}

Malware categorization was traditionally based on functionality, such as distinguishing between 
worms and viruses, or other types. However, modern malware often combines multiple capabilities, 
making purely functional classification difficult. A widely-used alternative approach is family-based 
classification, which groups malware according to shared codebases or common origins. This method 
enables tracking of malware evolution and facilitates more targeted defense mechanisms~\cite{aycock}.

The malware samples utilized in this study are derived from the RawMalTF dataset~\cite{rawmaltf}, which 
consists of malware samples collected from public repositories, including 
VirusShare~\cite{virusshare}, MalwareBazaar~\cite{bazaar}, and VXUnderground~\cite{VXU}. 
These samples are categorized by family labels, with~65 distinct malware families 
in total. We exclude families containing fewer than~1,000 samples, resulting in a set of~17 malware 
families. %From these families, various image types have been generated and analyzed in
The prior work by Agrawal, et al.~\cite{stamp2024malwareimages} evaluates 
eight image conversion techniques across these~17 malware families, providing one of the most 
comprehensive comparisons of binary-to-image transformation techniques considered to date.
We use this same malware image dataset for the research presented in this chapter.

The~17 malware families in our dataset are the following.
\begin{description}
\item[\textbf{\texttt{Agensla}}] scans system files and registry entries for stored credentials prior to exfiltration~\cite{agensla_kaspersky}.
\item[\textbf{\texttt{Androm}}] is a modular downloader/backdoor with anti-VM properties that fetches 
additional payloads~\cite{kasperskyandrom}.
\item[\textbf{\texttt{Convagent}}] is a Win32 Trojan-spy that intercepts keystrokes and generates screen 
captures~\cite{convagent}.
\item[\textbf{\texttt{Crypt}}] is a generic family for Microsoft Intermediate Language (MSIL)-based encryptors 
exploiting unpatched Windows systems~\cite{kasperskycrypt}.
\item[\textbf{\texttt{Crysan}}] is an MSIL backdoor that delivers modular stealer payloads~\cite{kasperskycrysan}.
\item[\textbf{\texttt{DCRat}}] is a .NET-based RAT with remote shell, file management, and webcam 
capabilities~\cite{kasperskydcrat}.
\item[\textbf{\texttt{Injuke}}] injects ransomware payloads to silently encrypt documents~\cite{kasperskyinjuke}.
\item[\textbf{\texttt{Makoob}}] is a cross-platform spyware that logs keystrokes and screenshots~\cite{kasperskymokes}.
\item[\textbf{\texttt{Mokes}}] is similar to Makoob, but primarily targeting macOS, although it has also 
been adapted for Windows and Android~\cite{kasperskymokes}.
\item[\textbf{\texttt{Noon}}] is a generic Trojan-spy that captures user activity and browser data~\cite{kasperskynoon}.
\item[\textbf{\texttt{Remcos}}] is a lightweight RAT designed for stealthy surveillance and remote 
control~\cite{kasperskyremcos}.
\item[\textbf{\texttt{Seraph}}] is a .NET-based credential stealer, specifically targeting browser and application 
credentials~\cite{kasperskyseraph}.
\item[\textbf{\texttt{SnakeLogger}}] is a modular keylogger that captures user input and session 
cookies~\cite{kasperskysnakelogger}.
\item[\textbf{\texttt{Stealerc}}] is an infostealer offered as malware-as-a-service, specializing in credential and 
crypto-wallet theft~\cite{kasperskystealerc}.
\item[\textbf{\texttt{Strab}}] is a Win32 Trojan capable of executing arbitrary commands on compromised 
systems~\cite{kasperskystrab}.
\item[\textbf{\texttt{Taskun}}] is an MSIL Trojan leveraging the Windows Task Scheduler for persistent 
execution~\cite{kasperskytaskun}.
\item[\textbf{\texttt{Zenpak}}] is a first-stage loader/backdoor associated with the BazarBackdoor 
family~\cite{kasperskyzenpak}.
\end{description}
For the experiments presented in this chapter, 1,000 samples have been selected from each of 
these~17 malware families, resulting in a dataset of~17,000 samples. As in~\cite{stamp2024malwareimages},
each malware sample is represented by its first~$224 \times 224 = 50{,}176$ bytes, which have been
transformed into eight distinct image-based representations for further analysis.

\subsection{Image Transformations}

\begin{figure}[!htb]
    \centering
     \begin{tabular}{|cc|cc|}\toprule
    \includegraphics[width=0.125\linewidth]{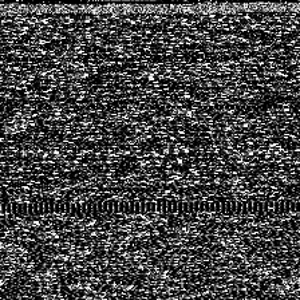}
    &
    \includegraphics[width=0.125\linewidth]{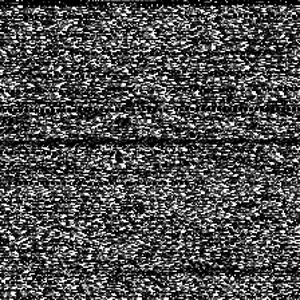}
    & %&
    \includegraphics[width=0.125\linewidth]{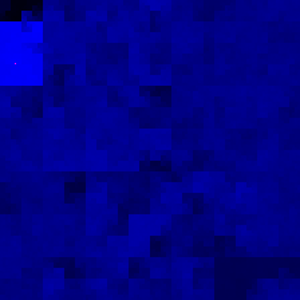}
    &
    \includegraphics[width=0.125\linewidth]{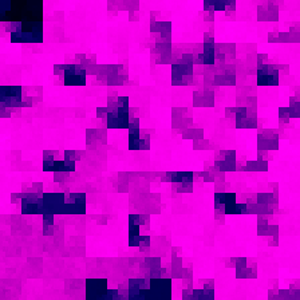}
    \\
    \adjustbox{scale=1.0}{\texttt{Agensla}}
    &
    \adjustbox{scale=1.0}{\texttt{Androm}}
    & %&
    \adjustbox{scale=1.0}{\texttt{Agensla}}
    &
    \adjustbox{scale=1.0}{\texttt{Androm}}
    \\
    \multicolumn{2}{|c|}{\adjustbox{scale=1.0}{(a) \grayscale}}
    &
    \multicolumn{2}{c|}{\adjustbox{scale=1.0}{(b) \entropyHilbert}}
    \\ \midrule
    \includegraphics[width=0.125\linewidth]{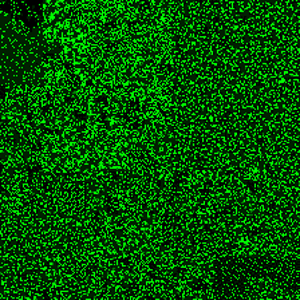}
    &
    \includegraphics[width=0.125\linewidth]{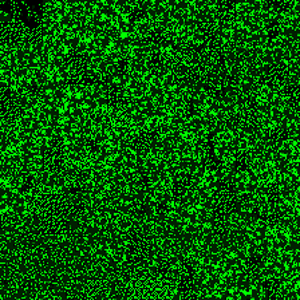}
    & %&
    \includegraphics[width=0.125\linewidth]{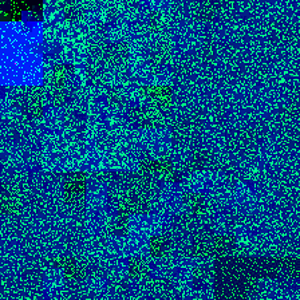}
    &
    \includegraphics[width=0.125\linewidth]{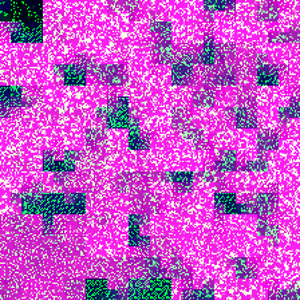}
    \\
    \adjustbox{scale=1.0}{\texttt{Agensla}}
    &
    \adjustbox{scale=1.0}{\texttt{Androm}}
    & %&
    \adjustbox{scale=1.0}{\texttt{Agensla}}
    &
    \adjustbox{scale=1.0}{\texttt{Androm}}
    \\
    \multicolumn{2}{|c|}{\adjustbox{scale=1.0}{(c) \byteclassHilbert}}
    &
    \multicolumn{2}{c|}{\adjustbox{scale=1.0}{(d) \hit}}
    \\ \midrule
    \includegraphics[width=0.125\linewidth]{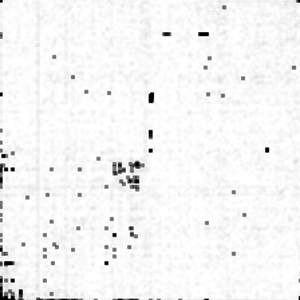}
    &
    \includegraphics[width=0.125\linewidth]{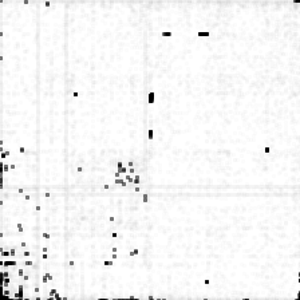}
    & %&
    \includegraphics[width=0.125\linewidth]{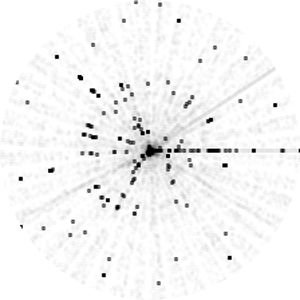}
    &
    \includegraphics[width=0.125\linewidth]{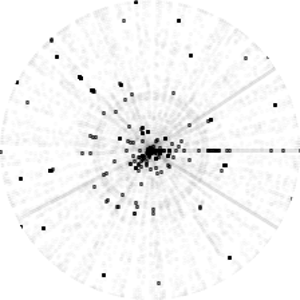}
    \\    
    \adjustbox{scale=1.0}{\texttt{Agensla}}
    &
    \adjustbox{scale=1.0}{\texttt{Androm}}
    & %&
    \adjustbox{scale=1.0}{\texttt{Agensla}}
    &
    \adjustbox{scale=1.0}{\texttt{Androm}}
    \\
    \multicolumn{2}{|c|}{\adjustbox{scale=1.0}{(e) \bigramCartesian}}
    &
    \multicolumn{2}{c|}{\adjustbox{scale=1.0}{(f) \bigramPolar}}
    \\ \midrule
    \includegraphics[width=0.125\linewidth]{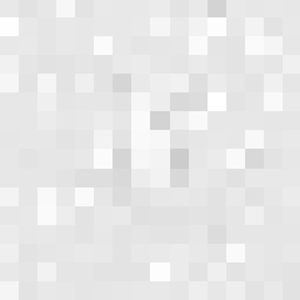}
    &
    \includegraphics[width=0.125\linewidth]{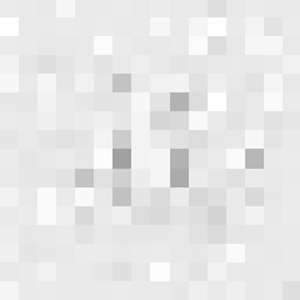}
    & %&
    \includegraphics[width=0.125\linewidth]{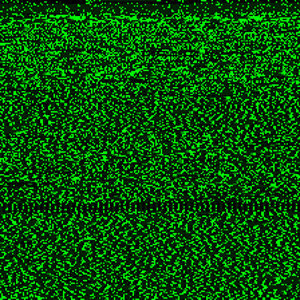}
    &
    \includegraphics[width=0.125\linewidth]{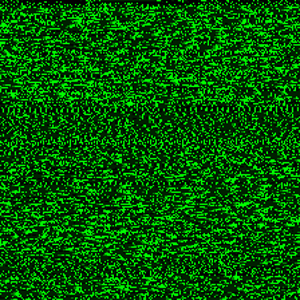}
    \\    
    \adjustbox{scale=1.0}{\texttt{Agensla}}
    &
    \adjustbox{scale=1.0}{\texttt{Androm}}
    & %&
    \adjustbox{scale=1.0}{\texttt{Agensla}}
    &
    \adjustbox{scale=1.0}{\texttt{Androm}}
    \\
    \multicolumn{2}{|c|}{\adjustbox{scale=1.0}{(g) \spiral}}
    &
    \multicolumn{2}{c|}{\adjustbox{scale=1.0}{(h) \byteclass}}
    \\ \bottomrule
    \end{tabular}
    \caption{Two examples of each image transformation type}\label{fig:all_sample}
\end{figure}

The eight image transformations that are applied to all of the~17,000 malware samples 
in our dataset are  illustrated in Figure~\ref{fig:all_sample}. Next, we briefly discuss each of these
image transformation techniques.

\subsubsection{\grayscale}

In the \grayscale\ transformation, raw bytes are directly mapped to pixel intensities. 
This is the simplest---and most commonly-used---approach, and serves as a baseline 
for comparison. This transformation often produces recognizable banding patterns and 
preserves byte-level texture information. A pair of examples of \grayscale\ images 
derived from malware executables appear in Figure~\ref{fig:all_sample}(a).

Note that the line breaks that are necessary
to create a 2-D grayscale image from an executable file result in artificial separations 
that are not present in the original binary. This may result in a loss of information when 
Convolutional Neural Networks (CNN) are used to classify such images, 
since CNNs only deal with local structure. Thus, it is reasonable to consider 
additional binary-to-image transformation techniques.

\subsubsection{\entropyHilbert\ Curve}

Entropy has long been used in malware analysis~\cite{BaysaLS13}. For example, entropy
can enable us to distinguish packed or encrypted regions from structured content within a binary. 
In the \entropyHilbert\ curve transformation~\cite{stamp2024malwareimages}, entropy values are 
computed over sliding windows of bytes and plotted along a Hilbert space-filling curve. 
The Hilbert curve layout preserves spatial locality better than a linear mapping. 
A pair of examples of \entropyHilbert\ curve images derived from malware executables appear 
in Figure~\ref{fig:all_sample}(b).

\subsubsection{\byteclassHilbert\ Curve}

Malware analysis involving character types, has proven successful~\cite{MIMURA2022100521}.
The \byteclassHilbert\ curve transformation~\cite{stamp2024malwareimages} maps each byte to a 
semantic category, such as printable characters, control characters, or instruction-like byte ranges, 
and plots these categories along a Hilbert curve. This grouping emphasizes code versus data segments 
and produces blocky textures that differ visually from raw byte representations. A pair of examples of 
\byteclassHilbert\ curve images derived from malware executables appear in Figure~\ref{fig:all_sample}(c).

\subsubsection{\hit}

The Hybrid Image Transformation (\hit)~\cite{stamp2024malwareimages} combines the entropy 
(red and blue channels) with \byteclass\ (green channel) into a single RGB image, using 
a Hilbert curve byte layout. Since the two component methods occupy different color channels, 
they can be merged without conflict, retaining hierarchical locality and producing distinctive geometric 
patterns. A pair of examples of \hit\ images derived from malware executables appear in 
Figure~\ref{fig:all_sample}(d).

\subsubsection{\bigramCartesian}

Byte bigrams---and more generally, $n$-grams---have been shown to be useful features
in malware analysis~\cite{TANG2023103118}.
The \bigramCartesian\ transformation~\cite{varela2017bigram} counts consecutive byte pairs 
and places their
relative frequencies into a~$256 \times 256$ matrix. The result is a frequency heatmap 
of byte transitions that captures statistical relationships between adjacent byte values.
A pair of examples of \bigramCartesian\ images derived from malware executables appear in 
Figure~\ref{fig:all_sample}(e).

\subsubsection{\bigramPolar}

The \bigramPolar\ transformation~\cite{varela2017bigram} uses the same bigram frequency counts as 
the \bigramCartesian\ method, but projects the result into a polar coordinate system. This can highlight 
radial symmetries and repeating byte patterns that are not easily detected in the Cartesian representation.
A pair of examples of \bigramPolar\ images derived from malware executables appear in Figure~\ref{fig:all_sample}(f).

\subsubsection{\spiral}

Gini information often appears in intrusion detection and malware research~\cite{Manokaran03042023}.
In the \spiral\ transformation~\cite{rawmaltf}, bytes are reordered using Gini importance 
scores derived from a Random Forest trained on byte histogram features. These Gini scores are plotted 
outward from the center of the image along a spiral path. This places the most informative bytes near 
the center of the image and produces a visually distinctive representation. A pair of examples of \spiral\ images 
derived from malware executables appear in Figure~\ref{fig:all_sample}(g).

\subsubsection{\byteclass}

Various types of semantic-based analysis have appeared in the malware literature~\cite{CHEN2026131781}.
The \byteclass\ transformation~\cite{stamp2024malwareimages} maps each byte to a categorical color 
value based on its ``semantic class,'' using a linear byte layout. This binary-to-image
transformation can be viewed as a lightweight alternative to the \byteclassHilbert\ curve transformation. 
Intuitively, this transformation
serves to highlight the high-level compositional structure of a binary by emphasizing the distribution 
of byte categories throughout the file. A pair of examples of \byteclass\ images derived from malware 
executables appear in Figure~\ref{fig:all_sample}(h).

\subsection{Grad-CAM and Explainable AI}

Gradient-weighted Class Activation Mapping (Grad-CAM) is a post-hoc explanation method for 
CNNs that highlights which regions in an input image contribute most to 
a model’s prediction. Conceptually, Grad-CAM computes the gradient of the class score with respect 
to the feature maps of a chosen convolutional layer, then aggregates these gradients to produce a 
class-specific importance weight for each feature map. A weighted sum of the feature maps is then 
passed through a ReLU activation to obtain a heatmap that emphasizes the spatial regions most 
responsible for the classification decision. When this heatmap is upsampled and overlaid on the original image, 
it provides an intuitive visualization of ``where the network is looking'' when it predicts a given class. 

Nazim et al.~\cite{grad_cam} apply Grad-CAM to malware imagery as part of a broader 
XAI framework that also includes SHAP and LIME. They use  Grad-CAM to 
generate heatmaps over byte-plot style images, showing which structural regions
drive the classifier’s decisions. The paper’s Grad-CAM examples demonstrate 
that the highlighted regions often align with semantically meaningful parts of the executable 
(e.g., header and code regions, resource sections), increasing trust in the model by 
linking visual attention patterns to domain knowledge about malware behavior. 

In addition to using Grad-CAM as a visualization tool, we also use the Grad-CAM heatmaps 
as input for further analysis. Grad-CAM maps are generated for each of our malware image 
representations. Features are then extracted from these heatmaps, and classical machine 
learning models are trained on the resulting feature vectors. For comparison, CNNs are also 
trained directly on the Grad-CAM images.
%In this way, Grad-CAM is used not just for explanation, but also as a new representation of the data. 
%The heatmaps both show what the original model focuses on and provide useful information 
%for classification tasks.

\section{Implementation}\label{chap:methodology}

In this chapter, a variety of classification experiments are considered.
We also present quantifiable XAI results.
In this section, we provide implementation details that are relevant
to the experimental results given in Section~\ref{chap:results}.

\subsection{Model Details}

All deep learning models in this chapter are implemented using TensorFlow/Keras. Classical machine learning models 
and evaluation utilities (e.g., accuracy, F1-score) are implemented using \texttt{scikit-learn}. 

Random Forest and eXtreme Gradient Boosting (XGBoost) are used as the 
classical classifiers throughout the experiments. 
%XGBoost is a scalable tree-boosting framework that combines multiple weak learners into a stronger 
%ensemble model while incorporating regularization and parallelized optimization strategies~\cite{chen2016xgboost}. 
Random Forest is an ensemble learning method that combines multiple decision trees to improve robustness and 
reduce variance~\cite{breiman2001random}, while XGBoost is a scalable gradient-boosting framework that
 combines weak learners into a stronger ensemble model using optimized tree-boosting strategies~\cite{chen2016xgboost}. 
 These models have been selected because they perform well on structured, high-dimensional feature representations 
 and can capture non-linear relationships between features.

Training is performed on an Apple Silicon Mac (M3~Pro), using mixed precision for 
CNNs to reduce training time. All experiments use fixed random seeds for reproducibility; in particular, 
$\mbox{\texttt{SEED}} = 42$ is set for Python's \texttt{random}, NumPy, and TensorFlow.
All CNN experiments use MobileNetV2 as the base, 
initialized with ImageNet weights. The MobileNetV2 architecture
was selected because it is lightweight and efficient to train on a laptop GPU.

Our overall CNN architecture consists of a MobileNetV2 backbone followed by a small classification head.
The MobileNetV2 backbone is implemented as
\begin{align*}
  & \mbox{\texttt{keras.applications.MobileNetV2(include\un top=False, }} \\
  & \hspace*{0.5in}\mbox{\texttt{weights="imagenet", input\un shape=(224, 224, 3))}}
\end{align*}
with \texttt{GlobalAveragePooling2D} applied to the final convolutional feature map.
The classification head is implemented with dropouts ($\mbox{dropout rate} = 0.3$),
a dense layer of size~256 with ReLU activation, and
$$
  \mbox{\texttt{dense}(\texttt{num\un classes}, $\mbox{\texttt{activation}} = \mbox{\texttt{softmax}}$)} 
$$
with \texttt{float32} output when using mixed precision.
Conceptually, the architecture can be viewed as
$$
  x \xrightarrow{\scalebox{0.85}{MobileNetV2}} f(x) \xrightarrow{\scalebox{0.85}{\texttt{GlobalAvgPool}}} 
  	z \xrightarrow{\scalebox{0.85}{$\mbox{Dropout}+\mbox{Dense}$}} h(z) \xrightarrow{\scalebox{0.85}{Dense}} \widehat{y}
$$
where~$\widehat{y}$ is a probability vector, the size of which
depends on the specific experiment.

Training is performed in two stages. In the first stage, the MobileNetV2 backbone is 
frozen ($\mbox{\texttt{base.trainable}} = \mbox{\texttt{False}}$) and only the classification head is trained 
using the Adam optimizer, with a learning rate of~$0.001$. In the second stage, 
the upper~30\%\ of the backbone layers (i.e., layers nearest the output) are unfrozen, 
excluding Batch Normalization (BatchNorm) layers, and the full model is fine-tuned using 
Adam with a reduced learning rate of~$0.0001$. 
 
In practice, the first training stage typically converges within approximately five to eight epochs, while the fine-tuning stage converges within approximately ten to twenty epochs depending on the transformation and validation performance. Early stopping and \texttt{ReduceLROnPlateau} are used in both stages to stabilize training and prevent overfitting.

All input images are represented using a fixed input size of~$224 \times 224$ for compatibility with MobileNetV2, 
and augmented during training with random horizontal flips, 
small random rotations, and small random zooms. On Apple Silicon, mixed precision (\texttt{mixed\un float16}) 
is enabled for speed, while the final softmax output is kept as \texttt{float32} for numerical stability.
For each experiment, stratified splits by family label are constructed using \texttt{StratifiedShuffleSplit}, 
allocating~70\%\ of samples for training, 15\%\ for validation, and~15\%\ for testing. This ensures that each 
malware family is proportionally represented across all three splits. For per-transformation experiments, 
the split is computed separately for each transformation.

\subsection{Grad-CAM}\label{sect:GC}

We use Grad-CAM to provide visual and quantifiable explanations, 
and also as a feature engineering technique. 
Grad-CAM produces a coarse heatmap that highlights regions most responsible 
for a particular class prediction~\cite{selvaraju2017gradcam}.

Let~$A^{k}$ denote the activation map of the last convolutional layer for channel~$k$, 
and let~$y^{c}$ denote the logit for class~$c$. Using \texttt{tf.GradientTape}, 
the gradients~$\partial y^{c} / \partial A^{k}_{ij}$ are computed and spatially averaged, 
that is
$$
  \alpha_{k}^{c} = \frac{1}{Z} \sum_{i,j} \frac{\partial y^{c}}{\partial A^{k}_{ij}}
$$
where~$(i, j)$ index spatial locations and~$Z$ is the number of spatial positions. 
The Grad-CAM map is then given by
$$
  L^{c}_{\text{Grad-CAM}} = \text{ReLU}\bigg(\sum_{k} \alpha_{k}^{c} A^{k}\bigg).
$$
The resulting map is normalized to~$[0, 1]$ and upsampled to the original image size using 
bilinear interpolation.
In practice, the model is split into a base (MobileNetV2) and a head (pooling and dense layers), 
and Grad-CAM is computed with respect to the last convolutional output of the base network. 
The true family label is used as the target class~$c$, 
so that the explanation corresponds to the ground-truth class.

A central contribution of this research is the construction of a comprehensive Grad-CAM 
dataset for all malware images under consideration. For each malware sample and image transformation type, three outputs are generated: the original transformed image, 
a raw CAM array (i.e., a normalized two-dimensional Grad-CAM heatmap), and an 
overlay image formed by blending the heatmap onto the original image using a colormap 
and alpha blending. The overlay visualization highlights the regions of the malware image 
that contribute most strongly to the CNN prediction.

\begin{figure}[!htb]
    \centering
%    \begin{tabular}{p{3cm}p{3cm}p{3cm}p{3cm}}
    \begin{tabular}{>{\centering\arraybackslash}m{0.2\linewidth} >{\centering\arraybackslash}m{0.2\linewidth} 
    	>{\centering\arraybackslash}m{0.2\linewidth} >{\centering\arraybackslash}m{0.2\linewidth}}
    \phantom{MMMMMMM} & \phantom{MMMMMMM} & \phantom{MMMMMMM} & \phantom{MMMMMMM} \\
    \multicolumn{4}{c}{\includegraphics[width=0.8\linewidth]{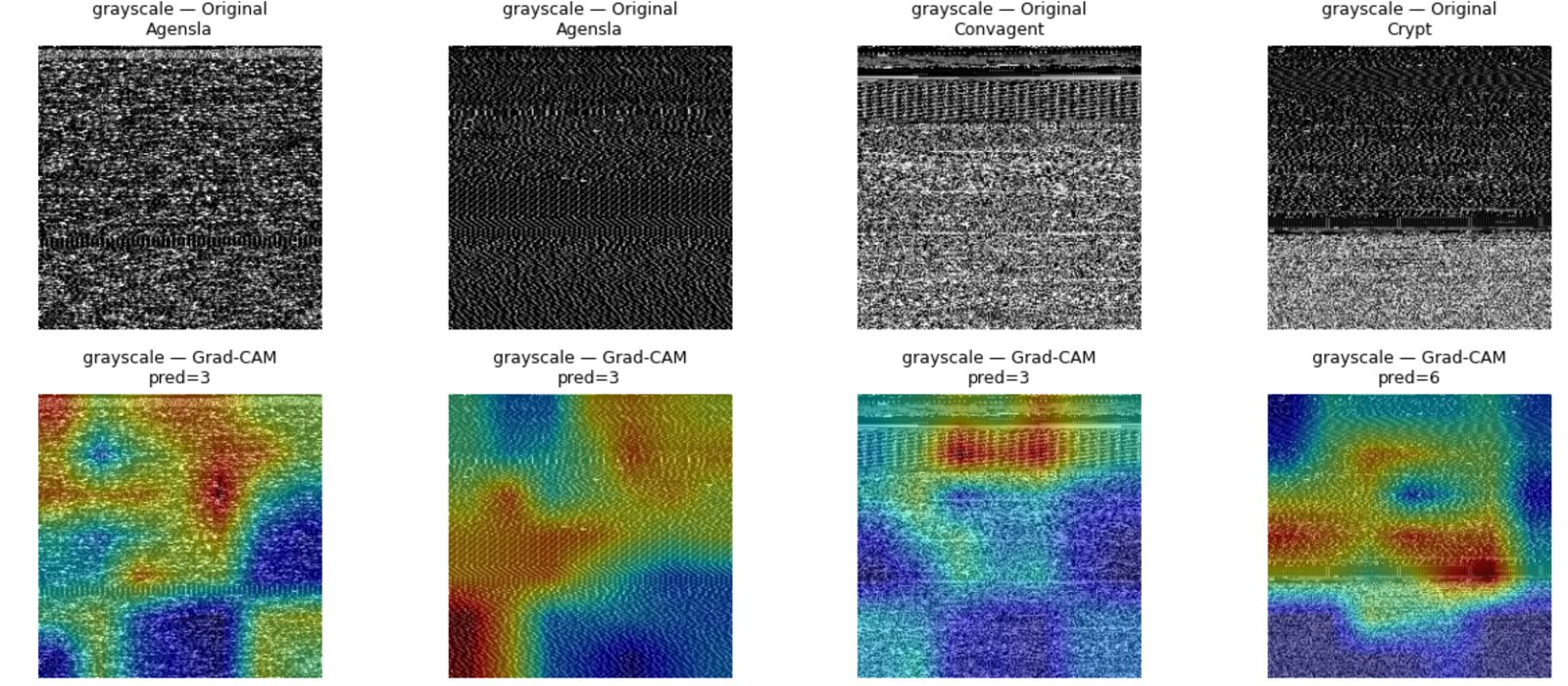}} \\[-1ex]
    \adjustbox{scale=1.0}{\ \ \ \ \texttt{Agensla}} & \adjustbox{scale=1.0}{\ \ \ \ \texttt{Agensla}} 
    	& \adjustbox{scale=1.0}{\ \ \texttt{Convagent}} & \adjustbox{scale=1.0}{\texttt{Crypt}} \\
    \multicolumn{4}{c}{\adjustbox{scale=1.0}{(a) \grayscale}} \\[2ex]
    \multicolumn{4}{c}{\includegraphics[width=0.8\linewidth]{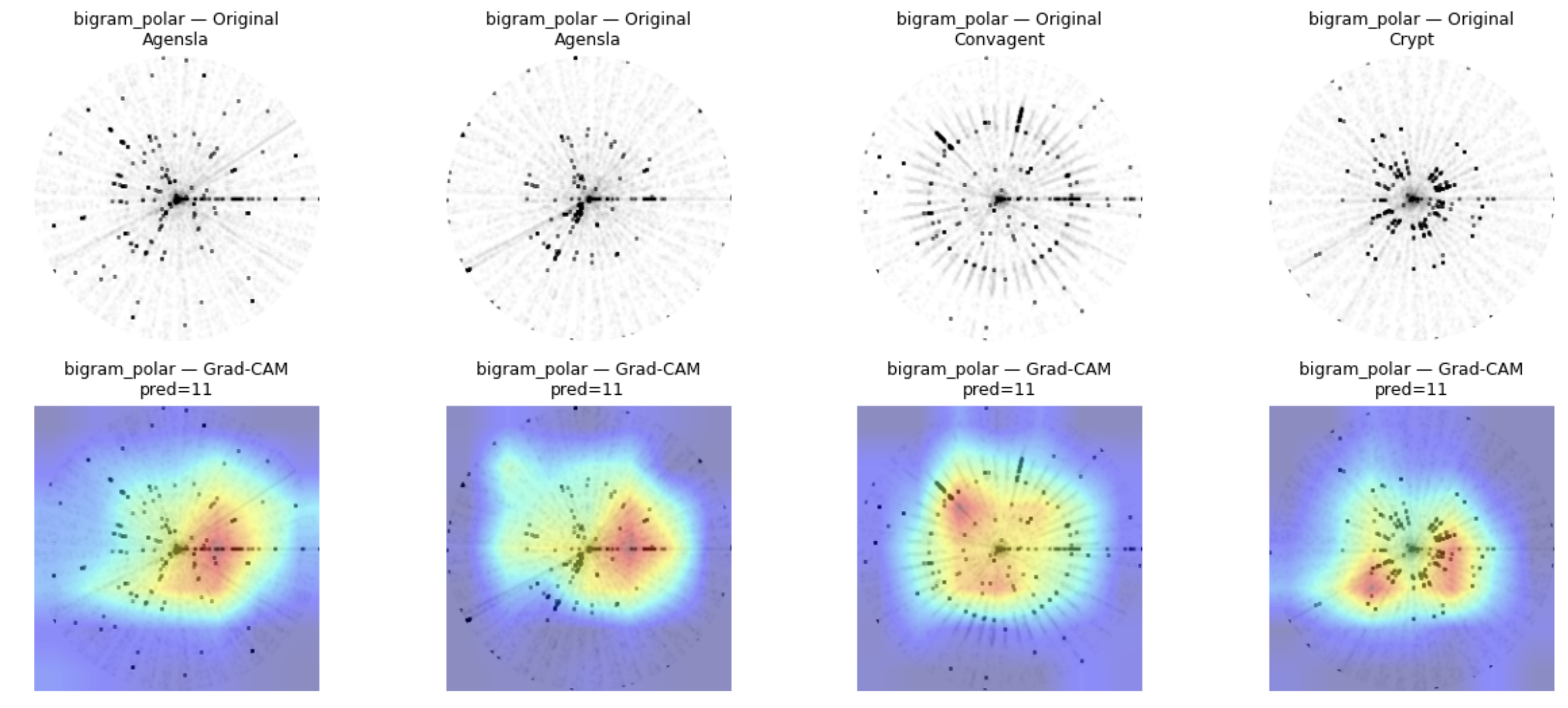}} \\[-1.5ex]
    \adjustbox{scale=1.0}{\ \ \ \ \texttt{Agensla}} & \adjustbox{scale=1.0}{\ \ \ \ \texttt{Agensla}} 
    	& \adjustbox{scale=1.0}{\ \ \texttt{Convagent}} & \adjustbox{scale=1.0}{\texttt{Crypt}} \\
    \multicolumn{4}{c}{\adjustbox{scale=1.0}{(b) \bigramPolar}}
    \end{tabular}
    \caption{Malware images and corresponding Grad-CAM overlays}\label{fig:gradcam}
\end{figure}

Figure~\ref{fig:gradcam} provides examples of malware images from two of the eight 
image conversion types considered, along with the corresponding Grad-CAM overlay 
images---examples for the other six image conversion types appear in Figures~\ref{fig:gradcam2a}
through~\ref{fig:gradcam2f} in the Appendix.
These Grad-CAM images highlight the spatial regions that the MobileNetV2-based 
classifier attends to when predicting each family label. These images visually demonstrate that 
family-specific structural patterns in the binaries translate into distinct, discriminative activation regions.

To quantitatively assess how much class information is contained in the Grad-CAM heatmaps---and 
inspired by traditional image-based malware classification---hand-crafted features are extracted from 
the CAMs and corresponding original images. %For each CAM, a feature vector is constructed by 
%combining intensity statistics, shape descriptors, and texture features.
From the normalized CAM, we compute the
mean and standard deviation of pixel values,
the~$25^{\thth}$ and~$75^{\thth}$ percentiles of the pixel intensity distribution,
and the fraction of pixels with intensity greater than~$0.5$ as a rough foreground ratio. 
These statistics capture whether the heatmap is globally focused or diffuse. 
A heatmap with low mean, high standard deviation, and a small foreground ratio indicates 
that attention is concentrated in a small region (focused), while a high mean with low standard 
deviation indicates broadly distributed attention (diffuse).

To quantify the geometric structure of the high-importance regions identified by Grad-CAM, 
shape-based features are extracted from thresholded CAM masks. 
The CAM is thresholded at~$0.5$ to obtain a binary mask of 
high-importance regions. From this mask, the connected components are extracted, 
and the largest blob is selected. Then the relative area of the largest blob, its eccentricity 
(via eigenvalues of the covariance matrix), and its solidity (ratio of area to convex hull area) 
are computed.
To capture transformation-invariant and family-invariant shape properties, Hu invariant moments 
are computed on the largest blob~\cite{hu_moment}. Hu moments are numerical descriptors that 
summarize the overall shape of the highlighted Grad-CAM regions. They are invariant to translation, 
scaling, and rotation, making them useful for capturing family-level structural patterns in the heatmaps.

Specifically, seven Hu moments are calculated and 
log-scaled for numerical stability as
$$
  \text{hu}_i = \hbox{}-\!\text{sign}(h_i) \log_{10}\big(|h_i| + \epsilon\big),
$$
where~$h_i$ are the seven classical Hu invariant moments defined by Hu~\cite{hu_moment} as 
algebraic combinations of normalized central moments up to order three, 
and~$\epsilon = 10^{-10}$ is used for numerical stability.

Local Binary Patterns (LBP) are computed on the normalized CAM~\cite{lbp}, 
using parameters~$P = 8$ neighbors, radius $R = 1$, and \texttt{method = "uniform"}. 
The uniform method restricts patterns to those with at most two bitwise transitions in 
the circular binary string, reducing the feature space to the most statistically common patterns. 
From the resulting LBP image, the mean, standard deviation, and histogram-derived summaries 
(such as entropy) are extracted. LBP features characterize local texture patterns in the high-activation regions.

Following prior feature-engineering approaches, Gray-Level Co-occurrence Matrices (GLCMs) are 
used to obtain rich texture statistics~\cite{glcm}. The procedure is as follows.
\begin{enumerate}
    \item Convert the original image to grayscale and normalize it to the interval~$[0,1]$.
    \item Use the CAM as a soft mask. Two thresholds are applied: 0.3 to emphasize 
    	low-to-medium attention regions and 0.7 to emphasize high-attention regions only.
    \item Quantize the image to 32 gray levels.
    \item Compute GLCMs with distances~$1$ and~$2$, and angles~$0$, $\pi/4$, $\pi/2$, and~$3\pi/4$, 
    covering horizontal, diagonal, vertical, and anti-diagonal spatial relationships. The symmetric 
    setting averages each direction with its opposite, and the normalized setting divides each 
    GLCM by its sum so that entries represent probabilities.
\end{enumerate}
From each GLCM, the following standard properties are computed.
\begin{description}
    \item[\textbf{Contrast}] measures local intensity variation as
    $$
    	\sum_{i,j} (i-j)^2\, p(i,j)
    $$
    \item[\textbf{Homogeneity}] measures closeness of elements to the diagonal as
    $$
        \sum_{i,j} \frac{p(i,j)}{1+|i-j|}
    $$
    \item[\textbf{Energy}] measures uniformity of the gray-level distribution as
    $$
        \sum_{i,j} p(i,j)^2
    $$ 
    \item[\textbf{Correlation}] measures linear dependency between neighboring pixels as
    $$
        \sum_{i,j} \frac{(i-\mu_i)(j-\mu_j)\,p(i,j)}{\sigma_i\,\sigma_j}
    $$
    where $\mu_i$ and $\mu_j$ are the mean values of the row and column marginal 
    distributions of $p(i,j)$, and $\sigma_i$ and $\sigma_j$ are their corresponding standard deviations.
\end{description}

For each property, the mean and standard deviation across all distances and angles are recorded, 
yielding eight values per property (mean and std at each of two thresholds). 
The feature names follow the pattern \texttt{low\_glcm\_contrast\_mean}, 
\texttt{low\_glcm\_contrast\_std}, \texttt{high\_glcm\_contrast\_mean}, 
\texttt{high\_glcm\_contrast\_std}, and analogously for homogeneity, energy, and correlation. 
These features capture texture patterns 
restricted to high-attention regions.

The combined feature vector per CAM %typically has on the order of 
has~50 dimensions. 
These~50 dimensions consist of~5 global intensity statistics (mean, standard deviation, 25th percentile, 
75th percentile, foreground ratio), 10 shape features (blob area, eccentricity, solidity, and 7 log-scaled Hu moments), 
3 LBP statistics (mean, standard deviation, entropy), and~32 GLCM statistics 
($\mbox{4 properties} \times \mbox{2 thresholds} \times \mbox{2 statistics} \times \mbox{2 distances}$, 
averaged across angles).

In Section~\ref{sect:ML_GC}, we give the results of experiments where
machine learning models are trained on these CAM-derived feature vectors to classify 
the~17 families in our dataset.
Stratified~5-fold cross-validation is used for each model, and  accuracy is reported. 

\subsection{HiResCAM}

As an alternative to Grad-CAM analysis, we consider HiResCAM~\cite{hirescam2020}. 
While Grad-CAM computes channel weights by globally average-pooling the gradients, 
then forming a weighted sum of activation maps, HiResCAM instead computes an element-wise 
product of the gradients and the activation maps before summing across channels. Formally, let~$A^k$ 
denote the activation map of the last convolutional layer for channel~$k$, 
and let~$\partial y^c\!/\partial A^k$ %~$\frac{\partial y^c}{\partial A^k}$ 
denote the gradient of the class score~$y^c$ 
with respect to that activation map. Then the HiResCAM map is given by
$$
  L^{c}_{\text{HiResCAM}} = \text{ReLU}\bigg(\sum_{k} \frac{\partial y^{c}}{\partial A^{k}} \odot A^{k}\bigg)
$$
where~``$\odot$'' denotes element-wise multiplication. This preserves spatial resolution in the gradient weighting, 
unlike Grad-CAM which collapses the spatial dimensions of the gradient via global averaging before weighting.
HiResCAM guarantees that positive evidence from any spatial location in the feature map 
contributes %positively 
to the final explanation, addressing a theoretical limitation of Grad-CAM.

We implement HiResCAM using the same \texttt{tf.GradientTape} pipeline as Grad-CAM. 
The same frozen MobileNetV2 backbone and head layers are used for both methods, 
ensuring a fair comparison. For each malware sample and transformation type, 
both a Grad-CAM heatmap and a HiResCAM heatmap are computed, normalized to~$[0,1]$, 
upsampled to~$224 \times 224$ via bilinear interpolation, and overlaid on the original image using 
the same \texttt{jet} colormap and~$\alpha = 0.45$ blending. We also compute
a pixel-level difference map
$$
    \left| L^{c}_{\text{Grad-CAM}} - L^{c}_{\text{HiResCAM}} \right|
$$ 
for each sample to quantify the difference between the two methods.

\begin{figure}[!htb]
    \centering
    \includegraphics[width=0.875\linewidth]{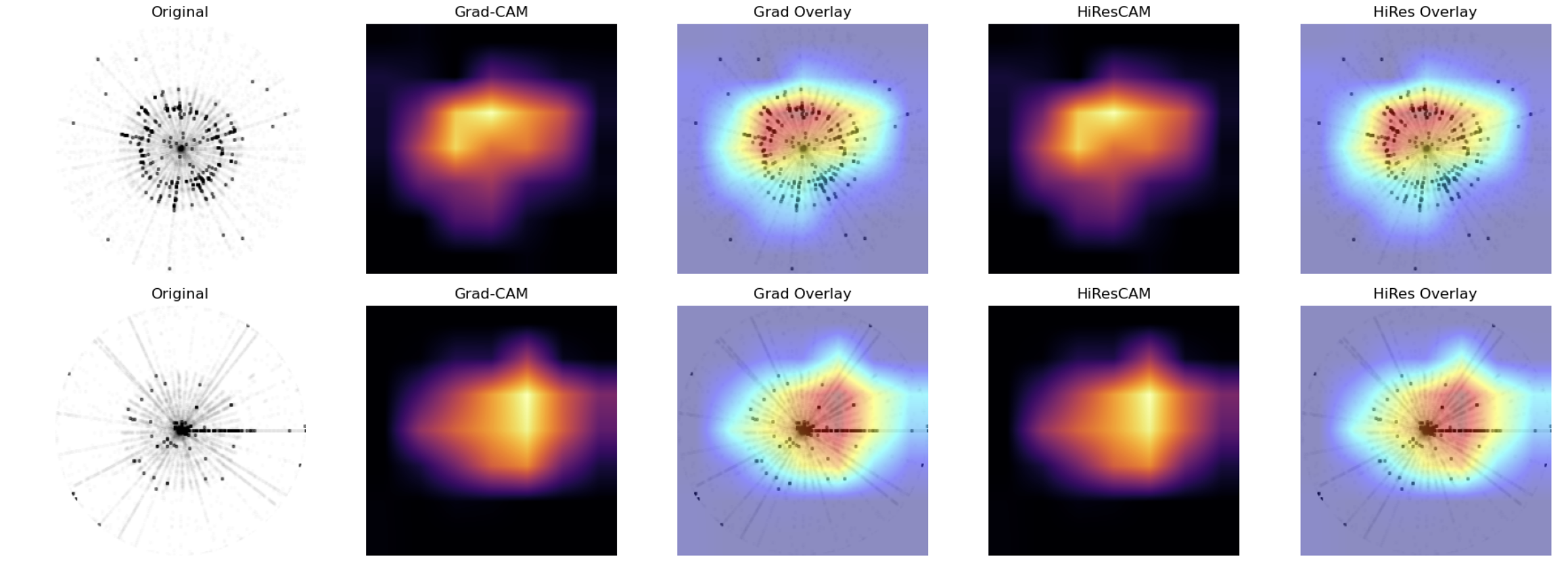}
    \caption{Comparison of Grad-CAM and HiResCAM heatmaps}
    \label{fig:hirescam}
\end{figure}

Figure~\ref{fig:hirescam} provides comparisons of Grad-CAM and HiResCAM 
heatmaps for two representative malware samples from the \bigramPolar\ transformation. 
Note that both the raw heatmaps and their overlaid versions are shown alongside the original images.
%We observe that when applied to the final convolutional layer of MobileNetV2, 
%Grad-CAM and HiResCAM produce virtually 
%identical heatmaps across all eight malware image transformations. The difference maps show localized 
%but subtle differences, primarily at the boundaries of high-activation regions. Qualitative inspection of multiple 
%samples from different malware families and transformations suggests that HiResCAM redistributes attention 
%at a finer spatial granularity even when the overall heatmap shape appears nearly identical to Grad-CAM.

The visual similarity between the Grad-CAM and HiResCAM results
in Figure~\ref{fig:hirescam} is striking.
Pixel-level difference maps between the two heatmaps show localized but small differences, primarily at the edges of activated regions, where the mean difference is $4 \times 10^{-8}$ and the maximum difference is $3 \times 10^{-7}$. This suggests that HiResCAM redistributes attention 
at a finer spatial granularity without substantially changing the overall explanation structure. 
However, these differences are negligible, and hence we only use Grad-CAM for
the quantitative evaluations  in Section~\ref{chap:results}.
This finding is consistent with the theoretical analysis in the HiResCAM paper~\cite{hirescam2020}, 
which notes that the Grad-CAM and HiResCAM tend to 
converge at the final convolutional layer when the spatial dimensions 
of the feature map are small---as is the case for MobileNetV2's final feature map, which is~$7 \times 7$. 
The element-wise formulation of HiResCAM would be expected to produce more 
distinct results at intermediate layers, where spatial resolution is higher. 
%Given the similarity of the two methods at the final layer and the focus of this 
%research on quantitative faithfulness and stability evaluation, Grad-CAM is selected used in
%all experiments reported in Section~\ref{chap:results}.

%Since each model is trained on its specific transformation's images, this experiment requires all eight 
%transform versions of each sample to be available. Each transform version is passed through its 
%corresponding model, and the resulting embeddings are concatenated. This means the 4096-dim 
%result represents the accuracy achievable when the full complement of transformation representations 
%is available, serving as an upper bound on combined-model performance. Features are extracted and 
%cached to disk one model at a time to avoid memory constraints.

\section{Experiments and Results}\label{chap:results}

In this section, we consider the performance of various models
when trained on different image types and features. %Among other related experiments, 
Specifically, we train models on each of the eight individual malware image transformations,
we consider the performance of 
CNN models trained on Grad-CAM overlay images, and
we train various hybrid models. These results show that we can improve on
the previous best results for our benchmark dataset.
In addition to our quest for improved accuracy, 
we provide quantitative XAI measures, based on Grad-CAM images.

\subsection{CNN Performance on Image Transformations}

Table~\ref{tab:rawcnn} gives the family classification accuracies for our~17-class dataset
when we train MobileNetV2 separately on each of the eight image transformations.
We observe that the entropy-based Hilbert curve images yield the strongest results, 
while \byteclassHilbert\ and \spiral\ give the weakest results. Furthermore, the differences between image types
are substantial, with the best offering a more than~50\%\ improvement over the worst.

\begin{table}[!htb]
\centering
\caption{CNN test accuracy on image transformations}\label{tab:rawcnn}
\begin{adjustbox}{scale=0.85}
\begin{tabular}{l c}
\toprule
\textbf{Transformation} & \textbf{Test accuracy} \\
\midrule
\grayscale              & 0.592 \\
\entropyHilbert        & \textbf{0.688} \\
\byteclassHilbert      & 0.433 \\
\hit                    & 0.532 \\
\bigramCartesian      & 0.601 \\
\bigramPolar          & 0.609 \\
\spiral                 & 0.489 \\
\byteclass              & 0.536 \\
\bottomrule
\end{tabular}
\end{adjustbox}
\end{table}

\subsection{Classical ML on Grad-CAM Features}\label{sect:ML_GC}

Next, we train several classical machine learning models on feature vectors extracted 
from Grad-CAM heatmaps, based on the 
features discussed in Section~\ref{sect:GC} 
(i.e., statistical, geometric, LBP, and GLCM features).
The results of these experiments are summarized 
in Table~\ref{tab:classical}.
While the numbers in Table~\ref{tab:classical} are not impressive, the
results all exceed the random accuracy of~$1/17=0.059$, 
indicating that the Grad-CAM heatmaps capture meaningful family-level information.

\begin{table}[!htb]
\centering
\caption{Classical models trained on Grad-CAM feature vectors}\label{tab:classical}
\begin{adjustbox}{scale=0.85}
\begin{tabular}{l c c}
\toprule
\textbf{Model} & \textbf{Accuracy} \\
\midrule
%%%%% What's the difference between Linear SVC and Linear SVM? Usually, SVC just refers to a linear SVM ---> I just used two different implementations of these, and yes it does mean the same thing. In hindsight, I should've just added one of these, so I have removed one of the results. 
Linear SVC       & $0.075$ \\ %\pm 0.014$ \\
SVM (RBF)       & $0.216$ \\ %\pm 0.014$ \\
XGBoost         & $0.275$ \\ %\pm 0.006$ \\
Random Forest   & \textbf{$0.294$} \\ % \pm 0.013$} \\
\bottomrule
\end{tabular}
\end{adjustbox}
\end{table}

\subsection{CNNs Trained on Grad-CAM Overlays}

In addition to training classical machine learning models on features extracted from 
Grad-CAM images, we also train CNNs directly on the Grad-CAM overlay images.
As above, we train one model per image type using the Grad-CAM overlay dataset.
The results of these experiments appear in Table~\ref{tab:pertransform}.
\entropyHilbert  and \grayscale\ transformations show the best performance,
with \spiral\ and \byteclassHilbert\ lagging far behind.

%For the next set of experiments, we train CNNs directly on 
%Grad-CAM overlay images instead of on the original transformed images.
%All overlay images across all transformations are pooled, with the label given by the corresponding
%malware family. A MobileNetV2-based CNN, with the same architecture as described above
%is trained on the overlay images. 

%To isolate the effect of each transformation, eight separate CNNs are trained, 
%one per transformation, using only the overlays corresponding to that transformation. 
%For each image transformation type, we train a MobileNetV2-based CNN to classify 
%the~17 families based on the overlay images.

\begin{table}[!htb]
\centering
\caption{CNN test accuracy per transformation (Grad-CAM overlays)}\label{tab:pertransform}
\begin{adjustbox}{scale=0.85}
\begin{tabular}{l c}
\toprule
\textbf{Transformation} & \textbf{Test accuracy} \\
\midrule
\grayscale             & 0.566 \\
\entropyHilbert       & \textbf{0.604} \\
\byteclassHilbert     & 0.433 \\
\hit                  & 0.532 \\
\bigramCartesian     & 0.555 \\
\bigramPolar         & 0.553 \\
\spiral                & 0.489 \\
\byteclass             & 0.536 \\
\bottomrule
\end{tabular}
\end{adjustbox}
\end{table}

\subsection{Hybrid CNN-HOG-XGBoost}

Motivated by prior strong performance of an XGBoost model trained on Histogram of Oriented Gradients (HOG) features
derived from \grayscale\ images~\cite{stamp2024malwareimages}, 
a hybrid pipeline is constructed. This hybrid model combines 
deep CNN embeddings derived from Grad-CAM overlays with classical HOG features, 
followed by an XGBoost classifier.

For each image transformation type, a CNN is first trained---as in the per-transformation experiments 
above---to classify the~17 families based on Grad-CAM overlays. Once trained, the CNN is used as a feature 
extractor: Each overlay image is passed through the network, and
the activations from the penultimate dense layer (a~256-dimensional vector) are extracted.
This 256-dimensional vector is used as the CNN embedding for the corresponding overlay.
From each overlay image, HOG~\cite{HOG} features are computed
by first converting the overlay to grayscale.
HOG descriptors are then extracted using 9 orientation bins, pixels-per-cell of $(8,8)$, 
cells-per-block of $(2,2)$, and $\ell_2$ block normalization, producing a medium-sized feature 
vector that captures local edge and texture structure. The resulting HOG feature vector 
is~$\ell_2$-normalized before concatenation with the CNN embedding. 
HOG emphasizes edge and structural patterns, complementing the higher-level 
semantic information captured by the CNN embeddings.

For each sample, the final feature vector is 
$$
  \text{feature} = (\text{CNN\un embedding} \,\|\, \text{HOG\un vector})
$$
that is, the feature vector is obtained by simply concatenating the CNN 
embedding and the HOG vector.
An XGBoost classifier is then trained as a~17-class model on these combined features.
%using a stratified train-validation-test split.

%Finally, we evaluate the hybrid pipeline in which CNN embeddings (256-dimensional) 
%are concatenated with HOG descriptors and classified using XGBoost.
The results of these experiments for each image transformation type 
are reported in Table~\ref{tab:hybrid}.
Here, \hit\ images perform best, with \byteclass, \byteclassHilbert, \grayscale, 
and \entropyHilbert\ all being competitive.
Overall, this hybrid approach consistently outperforms CNN-only models by
double-digit percentages, 
which is consistent with findings from previous work in~\cite{stamp2024malwareimages}, 
where handcrafted texture descriptors 
(e.g., HOG) are shown to enhance image-based malware classification.
% explanation for why the baseline accuracy is different between the two tables
Note that the CNN baseline accuracies in Table~\ref{tab:hybrid} differ slightly 
from those in Table~\ref{tab:pertransform} because the models were retrained 
independently in separate experimental runs using the same architecture, 
hyperparameters, and training procedure, with minor variation in results 
attributable to differences in random weight initialization between runs.

\begin{table}[!htb]
\centering
\caption{Hybrid model improvements (CNN vs CNN-HOG-XGBoost)}\label{tab:hybrid}
\begin{adjustbox}{scale=0.85}
\begin{tabular}{l c c}
\toprule
\multirow{2}{*}{\raisebox{-2pt}{\textbf{Transformation}}} 
	& \multicolumn{2}{c}{\textbf{Test accuracy}} \\ \cmidrule(lr){2-3}
  	& \textbf{CNN} & \textbf{Hybrid} \\
\midrule
\grayscale         & 0.565 & 0.717 \\
\entropyHilbert   & \textbf{0.630} & 0.711 \\
\byteclassHilbert         & 0.432 & 0.718 \\
\hit         & 0.552 & \textbf{0.731} \\
\bigramCartesian         & 0.581 & 0.658 \\
\bigramPolar         & 0.568 & 0.654 \\
\spiral         & 0.491 & 0.670 \\
\byteclass         & 0.501 & 0.721 \\
\bottomrule
\end{tabular}
\end{adjustbox}
\end{table}

\subsection{XAI Results}

To quantitatively measure whether Grad-CAM heatmaps correctly identify the image regions that drive model 
predictions, standard faithfulness metrics are computed. Specifically, we compute deletion AUC and 
insertion AUC, based on the perturbation-based evaluation framework 
described in~\cite{jei2023saliency, hama2023deletion, petsiuk2018rise}.

The deletion experiment progressively masks the highest-CAM pixels with a neutral baseline value 
(the image mean) and records the model's confidence at each step~\cite{jei2023saliency, hama2023deletion}. 
Formally, for a sorted ordering of pixels by CAM value, the image is modified at fractions~$f \in \{0.1, 0.2, \ldots, 1.0\}$ 
of total pixels. At each perturbation fraction~$f$, the modified image is passed back through the CNN, 
and the target-class softmax probability is recorded as~$p(f)$. The deletion AUC is computed as
$$
  \text{Deletion AUC} = \int_0^1 p(f)\, df \approx \frac{1}{|F|} \sum_{f \in F} p(f)
$$
A lower deletion AUC indicates that masking the high-CAM regions causes a larger confidence drop, 
meaning those regions were genuinely important to the prediction.

The insertion experiment starts from a blank baseline image and progressively reveals the 
highest-CAM pixels~\cite{petsiuk2018rise}. The insertion AUC is computed analogously to the
deletion AUC over the confidence values, but as more pixels are revealed. A higher insertion AUC 
indicates that the model can recover its confidence quickly from the high-CAM regions alone.

The faithfulness score is defined as
$$
  \text{faithfulness} = \text{insertion AUC} - \text{deletion AUC}
$$
Higher faithfulness values indicate that the CAM correctly identifies the pixels the model relies on. 
We also compute the following metrics.
\begin{itemize}
    \item drop\un top20pct --- Single-shot confidence drop when masking the top~20\% of CAM pixels.
    \item concentration\un p80 --- The fraction of total CAM mass contained in the top~20\% of pixels, 
    measuring how focused the heatmap is.
\end{itemize}
These faithfulness metrics are evaluated based on~200 samples per transformation, 
drawn via stratified random sampling from the full dataset.

Table~\ref{tab:faithfulness} contains faithfulness metrics results, averaged over 200 samples per
image transformation type. These results are consistent between the~50-sample 
and~200-sample evaluations, 
confirming that the rankings are stable and representative.

\begin{table}[!htb]
\centering
\caption{Faithfulness metrics per transformation (200 samples/transform)}
\label{tab:faithfulness}
\begin{adjustbox}{scale=0.75}
\begin{tabular}{l c c c c c}
\toprule
\multirow{2}{*}{\raisebox{-2pt}{\textbf{Transformation}}}
& \multicolumn{2}{c}{\textbf{AUC}}
& \multirow{2}{*}{\raisebox{-2pt}{\textbf{faithfulness}}}
& \multirow{2}{*}{\raisebox{-2pt}{\textbf{concentration\un p80}}}
& \multirow{2}{*}{\raisebox{-2pt}{\textbf{drop\un top20pct}}}
\\
\cmidrule(lr){2-3}
& \textbf{Deletion}
& \textbf{Insertion}
& 
& 
& 
\\
\midrule
\grayscale          & 0.075 & 0.131 & 0.056 & 0.443 & 0.122 \\
\entropyHilbert  & 0.090 & 0.211 & 0.121 & \textbf{0.588} & 0.376 \\
\byteclassHilbert & 0.102 & 0.155 & 0.053 & 0.432 & 0.087 \\
\hit                & 0.093 & 0.188 & 0.095 & 0.544 & 0.183 \\
\bigramCartesian & 0.075 & 0.182 & 0.106 & 0.455 & 0.211 \\
\bigramPolar    & 0.065 & 0.294 & \textbf{0.229} & 0.581 & \textbf{0.448} \\
\spiral             & 0.071 & 0.152 & 0.081 & 0.494 & 0.163 \\
\byteclass          & 0.059 & 0.176 & 0.118 & 0.486 & 0.272 \\
\bottomrule
\end{tabular}
\end{adjustbox}
\end{table}

From Table~\ref{tab:faithfulness}, we see that
\bigramPolar\ produces the most faithful Grad-CAM explanations despite being a mid-tier 
performer with respect to CNN accuracy. %, while \byteclassHilbert ranks last on faithfulness.
Also, \entropyHilbert\ ranks second on faithfulness while having the highest CNN accuracy, 
making it the most balanced transformation overall. 

%\subsection{Stability Results}

Stability measures whether Grad-CAM heatmaps produce consistent explanations under small 
perturbations of the input image. Stability is evaluated using a perturbation-based approach inspired 
by similarity-based robustness measures for saliency maps~\cite{chakraborty2022gradcam}. For each sample, 
Gaussian noise with standard deviation~$\sigma = 5$ (on a~0-255 pixel scale) is added to the original image, 
the CAM is regenerated using the same frozen model, and the perturbed CAM is compared to the saved 
reference CAM. This process is repeated~$n = 5$ times with independent noise draws, the results are averaged,
and the following four metrics are computed: Spearman correlation, mean-squared error (MSE),
structural similarity index measure (SSIM), top-$K$ overlap.

Spearman correlation is computed as a rank-order correlation between the original and 
perturbed CAM pixel values. This means that it measures whether the same pixels remain highly ranked in importance after perturbation. Higher values indicate the model identifies the same pixels 
as most important under noise.
MSE is based on the pixel-level differences between the original and perturbed CAMs. 
Lower values indicate more stable explanations.
SSIM captures whether the spatial layout and shape of the heatmap is preserved under noise.
Top-$K$ overlap is the fraction of the top~20\% most important pixels that are shared between the original 
and perturbed CAM. For example, a top-$K$ overlap score of~0.85 means~85\% of the highlighted 
pixels remain the same after adding noise.
These stability measures are all evaluated on~50 randomly-selected samples per transformation. 
Unlike faithfulness evaluation, which relies on perturbation curves with higher variability across samples, 
the stability metrics converged consistently with fewer samples, making 50 samples sufficient for 
comparative evaluation while reducing computational cost.

Table~\ref{tab:stability} contains stability metrics averaged over~50 samples per transformation,
with~$\sigma=5$ and~$n=5$ noise draws.
These stability rankings are nearly the inverse of the faithfulness rankings. For example,
\byteclass\ and \grayscale\ produce the most stable CAMs under noise, 
while \bigramPolar, which leads on faithfulness, ranks seventh on stability. 
We note that the \spiral\ image type is consistently the weakest 
transformation across both faithfulness and stability.

\begin{table}[!htb]
\centering
\caption{Stability metrics per transformation}
\label{tab:stability}
\begin{adjustbox}{scale=0.85}
\begin{tabular}{l c c c c}
\toprule
\textbf{Transformation} & \textbf{Spearman} & \textbf{MSE} & \textbf{SSIM} & \textbf{Top-K overlap} \\
\midrule
\grayscale          & 0.934 & 0.008 & 0.842 & 0.841 \\
\entropyHilbert    & 0.818 & 0.017 & 0.708 & 0.766 \\
\byteclassHilbert  & 0.917 & 0.010 & \textbf{0.872} & 0.832 \\
\hit                & 0.889 & 0.010 & 0.838 & 0.832 \\
\bigramCartesian  & 0.835 & 0.024 & 0.707 & 0.729 \\
\bigramPolar      & 0.687 & 0.040 & 0.485 & 0.658 \\
\spiral             & 0.547 & 0.061 & 0.474 & 0.513 \\
\byteclass          & \textbf{0.944} & \textbf{0.007} & 0.863 & \textbf{0.857} \\
\bottomrule
\end{tabular}
\end{adjustbox}
\end{table}

%\subsubsection{Faithfulness vs Stability Tradeoff}

A key finding is that no single transformation dominates both faithfulness and stability. In fact,
transformations that produce highly faithful explanations tend to have lower stability, and vice versa. 
This tradeoff reflects a fundamental tension between explanation correctness and explanation robustness.
Specifically, since \bigramPolar\ is first in faithfulness, but seventh in stability, models trained on
this image type identify genuinely important pixels, but the specific localization is sensitive to 
small input changes. On the other hand, \byteclass\ and \grayscale\ are the most stable, 
but weak with respect to faithfulness, implying that Grad-CAM images are relatively consistent, 
but do not necessarily point to the most predictive regions. In some sense, the best overall 
balance may be provided by \entropyHilbert, since it has the highest CNN accuracy (0.688), 
second-highest faithfulness (0.121), and moderate stability (0.818).

%\subsection{Confusion Matrix Comparison}

To directly compare whether Grad-CAM overlay images convey different or similar classification information 
to the original images, separate MobileNetV2-based CNNs are trained on each, per transformation type. 
The same two-stage training procedure is used, with the same 70:15:15 stratified train:validation:test
split for both image types. For each transformation, we generate a confusion matrix
for the model trained in the original images and a confusion matrix for the model trained on the
overlay images. To quantify similarity between these two confusion matrices, Spearman rank correlation 
is computed between their flattened vectors. A value close to~1.0 indicates that the two image types 
produce nearly identical classification errors, while lower values indicate meaningfully different behavior.

Table~\ref{tab:cmcomparison} summarizes per-transformation CNN accuracy on 
the original images and Grad-CAM overlay images, along with the Spearman similarity 
between the corresponding confusion matrices.

\begin{table}[!htb]
\centering
\caption{CNN accuracy and confusion matrix similarity} % (original vs overlay images)}
\label{tab:cmcomparison}
\begin{adjustbox}{scale=0.85}
\begin{tabular}{l c c c c}
\toprule
%\multirow{2}{*}{\raisebox{-3pt}{\textbf{Transformation}}} & \multicolumn{2}{c}{\textbf{Accuracy}} 
%	& \multirow{2}{*}{\raisebox{-3pt}{\textbf{Difference}}} 
%	& \multirow{2}{*}{\raisebox{-3pt}{\textbf{CM Sim}}} \\ \cmidrule(lr){2-3}
%  	& \textbf{Original} & \textbf{Overlay}  \\
\multirow{2}{*}{\raisebox{-2pt}{\textbf{Transformation}}} & \multicolumn{2}{c}{\textbf{Accuracy}} 
	& \multirow{2}{*}{\raisebox{-2pt}{\textbf{Difference}}} 
	& \raisebox{-3pt}{\textbf{Confusion matrix}} \\ \cmidrule(lr){2-3}
  	& \textbf{Original} & \textbf{Overlay} & & \raisebox{2pt}{\textbf{similarity}} \\
\midrule
\grayscale          & 0.542 & 0.602 & $\mathbf{\hbox{}+\!0.060}$ & 0.832 \\
\entropyHilbert    & \textbf{0.628} & \textbf{0.624} & $\hbox{}-\!0.004$ & 0.836 \\
\byteclassHilbert  & 0.348 & 0.400 & $\hbox{}+\!0.053$ & 0.662 \\
\hit                & 0.551 & 0.548 & $\hbox{}-\!0.003$ & \textbf{0.859} \\
\bigramCartesian  & 0.587 & 0.566 & $\hbox{}-\!0.021$ & 0.797 \\
\bigramPolar      & 0.566 & 0.538 & $\hbox{}-\!0.028$ & 0.841 \\
\spiral             & 0.481 & 0.499 & $\hbox{}+\!0.018$ & 0.819 \\
\byteclass          & 0.537 & 0.572 & $\hbox{}+\!0.035$ & 0.793 \\
\bottomrule
\end{tabular}
\end{adjustbox}
\end{table}

For five out of the eight image transformation types, overlay images perform equal to or better than 
the original images, indicating that Grad-CAM overlays generally preserve---and sometimes 
enhance---family-discriminative structure. The largest improvements when using the overlay
images are for
\grayscale~\hbox{(\!\!$\hbox{}+\!6.0\%$)} and \byteclassHilbert~\hbox{(\!\!$\hbox{}+\!5.3\%$)}.
Bigram-based transformations are the only cases where original images outperform overlays, 
suggesting that the CAM overlay may obscure the frequency-domain structure that these 
representations encode.

Confusion matrices for models trained on \grayscale\ and \byteclassHilbert\ images
(along with the corresponding overlay images) are provided in Figures~\ref{fig:conf}(a) 
through~\ref{fig:conf}(d). 
%Confusion matrix similarity scores range from 0.662 (\byteclassHilbert) to 0.859 (\hit). 
%\byteclassHilbert shows the most divergent error behavior between image types, 
%making it a candidate for further investigation. 
Per-family analysis reveals that \texttt{Noon}, \texttt{Androm}, \texttt{Agensla}, and \texttt{Injuke} are 
among the bottom five 
families (by accuracy) in nearly every experiment, suggesting that these families have 
binary structures that are difficult to distinguish, regardless of the image representation.

%\begin{figure}[!htb]
%    \centering
%    \includegraphics[width=0.85\linewidth]{images/grayscale_confmatrix.png}
%    \caption{Confusion matrix for grayscale transformation}
%    \label{fig:grayscaleconf}
%\end{figure}

%\begin{figure}[!htb]
%    \centering
%    \includegraphics[width=0.85\linewidth]{images/byte_hcurve_confmatrix.png}
%    \caption{Confusion matrix for \byteclassHilbert transformation}
%    \label{fig:bclassconf}
%\end{figure}

% \begin{figure}[H]
%     \centering
%     \begin{tabular}{cc}
%     \includegraphics[scale=0.365]{images/conf_gray_orig.png}
%     &
%     \includegraphics[scale=0.365]{images/conf_gray_overlay.png}
%     \\
%     \adjustbox{scale=1.0}{(a) Grayscale (original images)}
%     &
%     \adjustbox{scale=1.0}{(b) Grayscale (overlay images)}
%     \\[2ex]
%     \includegraphics[scale=0.365]{images/conf_bch_orig.png}
%     &
%     \includegraphics[scale=0.365]{images/conf_bch_overlay.png}
%     \\
%     \adjustbox{scale=1.0}{(c) Byteclass Hilbert (original images)}
%     &
%     \adjustbox{scale=1.0}{(d) Byteclass Hilbert (overlay images)}
%     \end{tabular}
%     \caption{Confusion matrices}\label{fig:conf}
% \end{figure}

\begin{figure}[!htb]
    \centering
    \advance\tabcolsep by -8.0pt
    \begin{tabular}{cc}
    \scalebox{0.9}{%
        % Grayscale — Original Images (acc=0.542)
\begin{tikzpicture}[scale=1.0]
    \begin{axis}[
        width=7.0cm, height=7.0cm,
        colormap={bluewhite}{color=(white) rgb255=(100,149,237)},
        xticklabels={Agensla,Androm,Convagent,Crypt,Crysan,DCRat,Injuke,Makoob,Mokes,Noon,Remcos,Seraph,SnakeLogger,Stealerc,Strab,Taskun,Zenpak},
        xtick={0,...,16}, xtick style={draw=none},
        xticklabel style={scale=0.85,anchor=east,rotate=60,yshift=-5pt,font=\tt},
        yticklabels={Agensla,Androm,Convagent,Crypt,Crysan,DCRat,Injuke,Makoob,Mokes,Noon,Remcos,Seraph,SnakeLogger,Stealerc,Strab,Taskun,Zenpak},
        ytick={0,...,16}, ytick style={draw=none},
        enlargelimits=false,
        yticklabel style={scale=0.85,font=\tt},
        colorbar, colorbar style={
            colorbar/width=2mm,
            ytick={0,30,60,90,120,150}, yticklabels={0,30,60,90,120,150},
            yticklabel={\pgfmathprintnumber\tick},
            yticklabel style={scale=0.9,/pgf/number format/fixed,/pgf/number format/fixed zerofill,/pgf/number format/precision=0}},
        colorbar=false,
        point meta min=0, point meta max=150,
        nodes near coords={\pgfmathprintnumber\pgfplotspointmeta},
        nodes near coords black white/.style={
            small value/.style={yshift=-5pt,text=black,/pgf/number format/fixed,/pgf/number format/precision=0,/pgf/number format/zerofill=true,scale=0.6},
            large value/.style={yshift=-5pt,text=white,/pgf/number format/fixed,/pgf/number format/precision=0,/pgf/number format/zerofill=true,scale=0.6},
            every node near coord/.style={check for zero/.code={
                \pgfmathfloatifflags{\pgfplotspointmeta}{0}{\pgfkeys{/tikz/coordinate}}{
                    \begingroup\pgfkeys{/pgf/fpu}\pgfmathparse{\pgfplotspointmeta<#1}
                    \global\let\result=\pgfmathresult\endgroup
                    \pgfmathfloatcreate{1}{1.0}{0}\let\ONE=\pgfmathresult
                    \ifx\result\ONE\pgfkeysalso{/pgfplots/small value}\else\pgfkeysalso{/pgfplots/large value}\fi}},check for zero}},
        nodes near coords black white=75,
    ]
\addplot[matrix plot, mesh/cols=17, point meta=explicit, draw=gray]
table [x=x, y=y, meta=C] {
x y C
0 0 64
1 0 4
2 0 0
3 0 5
4 0 1
5 0 0
6 0 1
7 0 0
8 0 0
9 0 18
10 0 8
11 0 14
12 0 5
13 0 4
14 0 2
15 0 24
16 0 0
0 1 14
1 1 50
2 1 0
3 1 6
4 1 0
5 1 0
6 1 2
7 1 12
8 1 1
9 1 17
10 1 3
11 1 11
12 1 4
13 1 15
14 1 5
15 1 9
16 1 1
0 2 1
1 2 1
2 2 98
3 2 1
4 2 0
5 2 0
6 2 0
7 2 15
8 2 0
9 2 2
10 2 1
11 2 1
12 2 0
13 2 13
14 2 6
15 2 1
16 2 10
0 3 30
1 3 5
2 3 0
3 3 61
4 3 3
5 3 0
6 3 5
7 3 0
8 3 0
9 3 4
10 3 2
11 3 15
12 3 5
13 3 13
14 3 0
15 3 6
16 3 1
0 4 11
1 4 0
2 4 1
3 4 3
4 4 107
5 4 0
6 4 4
7 4 0
8 4 0
9 4 2
10 4 1
11 4 8
12 4 2
13 4 9
14 4 1
15 4 0
16 4 1
0 5 0
1 5 0
2 5 0
3 5 1
4 5 2
5 5 133
6 5 0
7 5 0
8 5 0
9 5 0
10 5 2
11 5 2
12 5 0
13 5 8
14 5 2
15 5 0
16 5 0
0 6 12
1 6 0
2 6 9
3 6 5
4 6 5
5 6 0
6 6 50
7 6 0
8 6 1
9 6 7
10 6 6
11 6 16
12 6 1
13 6 28
14 6 6
15 6 4
16 6 0
0 7 0
1 7 13
2 7 0
3 7 0
4 7 0
5 7 0
6 7 1
7 7 120
8 7 0
9 7 0
10 7 0
11 7 2
12 7 1
13 7 0
14 7 13
15 7 0
16 7 0
0 8 0
1 8 2
2 8 2
3 8 0
4 8 0
5 8 0
6 8 2
7 8 0
8 8 42
9 8 0
10 8 0
11 8 1
12 8 0
13 8 35
14 8 13
15 8 0
16 8 53
0 9 19
1 9 7
2 9 1
3 9 2
4 9 2
5 9 0
6 9 4
7 9 0
8 9 0
9 9 44
10 9 11
11 9 12
12 9 8
13 9 6
14 9 2
15 9 32
16 9 0
0 10 12
1 10 4
2 10 0
3 10 4
4 10 3
5 10 0
6 10 4
7 10 2
8 10 0
9 10 11
10 10 79
11 10 11
12 10 2
13 10 6
14 10 5
15 10 7
16 10 0
0 11 17
1 11 3
2 11 0
3 11 3
4 11 4
5 11 0
6 11 3
7 11 0
8 11 0
9 11 0
10 11 2
11 11 106
12 11 3
13 11 5
14 11 3
15 11 0
16 11 1
0 12 11
1 12 4
2 12 0
3 12 6
4 12 2
5 12 0
6 12 1
7 12 0
8 12 1
9 12 5
10 12 8
11 12 6
12 12 89
13 12 1
14 12 3
15 12 13
16 12 0
0 13 2
1 13 3
2 13 8
3 13 5
4 13 1
5 13 0
6 13 2
7 13 2
8 13 5
9 13 0
10 13 4
11 13 5
12 13 0
13 13 90
14 13 14
15 13 0
16 13 9
0 14 0
1 14 0
2 14 1
3 14 0
4 14 0
5 14 0
6 14 0
7 14 2
8 14 3
9 14 0
10 14 20
11 14 0
12 14 0
13 14 36
14 14 88
15 14 0
16 14 0
0 15 26
1 15 9
2 15 0
3 15 0
4 15 0
5 15 0
6 15 1
7 15 0
8 15 0
9 15 26
10 15 4
11 15 5
12 15 7
13 15 3
14 15 0
15 15 69
16 15 0
0 16 0
1 16 3
2 16 3
3 16 0
4 16 0
5 16 0
6 16 0
7 16 0
8 16 19
9 16 0
10 16 1
11 16 3
12 16 0
13 16 21
14 16 9
15 16 0
16 16 91
};
\end{axis}
\end{tikzpicture}
    }
    &
    \scalebox{0.9}{%
        % Grayscale — Grad-CAM Overlay Images (acc=0.602)
\begin{tikzpicture}[scale=1.0]
    \begin{axis}[
        width=7.0cm, height=7.0cm,
        colormap={bluewhite}{color=(white) rgb255=(100,149,237)},
        xticklabels={Agensla,Androm,Convagent,Crypt,Crysan,DCRat,Injuke,Makoob,Mokes,Noon,Remcos,Seraph,SnakeLogger,Stealerc,Strab,Taskun,Zenpak},
        xtick={0,...,16}, xtick style={draw=none},
        xticklabel style={scale=0.85,anchor=east,rotate=60,yshift=-5pt,font=\tt},
        yticklabels={Agensla,Androm,Convagent,Crypt,Crysan,DCRat,Injuke,Makoob,Mokes,Noon,Remcos,Seraph,SnakeLogger,Stealerc,Strab,Taskun,Zenpak},
        ytick={0,...,16}, ytick style={draw=none},
        enlargelimits=false,
        yticklabel style={scale=0.85,font=\tt},
        colorbar, colorbar style={
            colorbar/width=2mm,
            ytick={0,30,60,90,120,150}, yticklabels={0,30,60,90,120,150},
            yticklabel={\pgfmathprintnumber\tick},
            yticklabel style={scale=0.825,/pgf/number format/fixed,/pgf/number format/fixed zerofill,/pgf/number format/precision=0}},
        point meta min=0, point meta max=150,
        nodes near coords={\pgfmathprintnumber\pgfplotspointmeta},
        nodes near coords black white/.style={
            small value/.style={yshift=-5pt,text=black,/pgf/number format/fixed,/pgf/number format/precision=0,/pgf/number format/zerofill=true,scale=0.6},
            large value/.style={yshift=-5pt,text=white,/pgf/number format/fixed,/pgf/number format/precision=0,/pgf/number format/zerofill=true,scale=0.6},
            every node near coord/.style={check for zero/.code={
                \pgfmathfloatifflags{\pgfplotspointmeta}{0}{\pgfkeys{/tikz/coordinate}}{
                    \begingroup\pgfkeys{/pgf/fpu}\pgfmathparse{\pgfplotspointmeta<#1}
                    \global\let\result=\pgfmathresult\endgroup
                    \pgfmathfloatcreate{1}{1.0}{0}\let\ONE=\pgfmathresult
                    \ifx\result\ONE\pgfkeysalso{/pgfplots/small value}\else\pgfkeysalso{/pgfplots/large value}\fi}},check for zero}},
        nodes near coords black white=75,
    ]
\addplot[matrix plot, mesh/cols=17, point meta=explicit, draw=gray]
table [x=x, y=y, meta=C] {
x y C
0 0 71
1 0 2
2 0 1
3 0 7
4 0 3
5 0 0
6 0 7
7 0 0
8 0 0
9 0 21
10 0 1
11 0 8
12 0 7
13 0 3
14 0 0
15 0 19
16 0 0
0 1 15
1 1 69
2 1 1
3 1 6
4 1 2
5 1 0
6 1 9
7 1 11
8 1 2
9 1 3
10 1 1
11 1 6
12 1 2
13 1 12
14 1 5
15 1 5
16 1 1
0 2 1
1 2 1
2 2 91
3 2 0
4 2 1
5 2 0
6 2 17
7 2 0
8 2 3
9 2 2
10 2 1
11 2 0
12 2 1
13 2 18
14 2 5
15 2 1
16 2 8
0 3 17
1 3 4
2 3 0
3 3 78
4 3 1
5 3 0
6 3 8
7 3 0
8 3 0
9 3 10
10 3 3
11 3 14
12 3 1
13 3 6
14 3 1
15 3 6
16 3 1
0 4 11
1 4 2
2 4 1
3 4 3
4 4 99
5 4 1
6 4 6
7 4 0
8 4 0
9 4 7
10 4 2
11 4 9
12 4 1
13 4 6
14 4 1
15 4 1
16 4 0
0 5 0
1 5 0
2 5 0
3 5 2
4 5 2
5 5 138
6 5 0
7 5 0
8 5 0
9 5 1
10 5 2
11 5 0
12 5 0
13 5 2
14 5 3
15 5 0
16 5 0
0 6 5
1 6 1
2 6 10
3 6 2
4 6 6
5 6 1
6 6 65
7 6 0
8 6 2
9 6 9
10 6 3
11 6 12
12 6 2
13 6 24
14 6 5
15 6 3
16 6 0
0 7 0
1 7 6
2 7 0
3 7 0
4 7 0
5 7 0
6 7 0
7 7 137
8 7 0
9 7 0
10 7 0
11 7 0
12 7 0
13 7 0
14 7 7
15 7 0
16 7 0
0 8 0
1 8 2
2 8 0
3 8 0
4 8 0
5 8 1
6 8 0
7 8 0
8 8 91
9 8 0
10 8 0
11 8 0
12 8 0
13 8 28
14 8 2
15 8 0
16 8 26
0 9 35
1 9 7
2 9 0
3 9 14
4 9 0
5 9 0
6 9 4
7 9 0
8 9 0
9 9 54
10 9 6
11 9 7
12 9 3
13 9 3
14 9 2
15 9 15
16 9 0
0 10 10
1 10 4
2 10 1
3 10 2
4 10 2
5 10 1
6 10 5
7 10 2
8 10 0
9 10 11
10 10 86
11 10 6
12 10 1
13 10 7
14 10 3
15 10 9
16 10 0
0 11 11
1 11 5
2 11 0
3 11 6
4 11 6
5 11 1
6 11 14
7 11 0
8 11 0
9 11 1
10 11 1
11 11 89
12 11 5
13 11 4
14 11 1
15 11 5
16 11 1
0 12 19
1 12 0
2 12 0
3 12 0
4 12 3
5 12 0
6 12 4
7 12 0
8 12 1
9 12 6
10 12 2
11 12 9
12 12 100
13 12 0
14 12 3
15 12 3
16 12 0
0 13 1
1 13 0
2 13 3
3 13 7
4 13 3
5 13 0
6 13 8
7 13 1
8 13 11
9 13 2
10 13 4
11 13 5
12 13 0
13 13 90
14 13 13
15 13 0
16 13 2
0 14 0
1 14 1
2 14 0
3 14 0
4 14 0
5 14 0
6 14 0
7 14 7
8 14 4
9 14 0
10 14 1
11 14 0
12 14 0
13 14 14
14 14 118
15 14 0
16 14 5
0 15 37
1 15 3
2 15 0
3 15 5
4 15 5
5 15 1
6 15 1
7 15 0
8 15 0
9 15 17
10 15 3
11 15 1
12 15 2
13 15 1
14 15 1
15 15 73
16 15 0
0 16 0
1 16 0
2 16 1
3 16 0
4 16 0
5 16 0
6 16 2
7 16 0
8 16 36
9 16 0
10 16 0
11 16 0
12 16 0
13 16 21
14 16 4
15 16 0
16 16 86
};
\end{axis}
\end{tikzpicture}%
    }
    \\
    \adjustbox{scale=0.85}{(a) \grayscale\ (original images)}
    &
    \adjustbox{scale=0.85}{(b) \grayscale\ (overlay images)}
    \\[2ex]
    \scalebox{0.9}{%
        % byteclass_hcurve — Original Images (acc=0.348)
\begin{tikzpicture}[scale=1.0]
    \begin{axis}[
        width=7.0cm, height=7.0cm,
        colormap={bluewhite}{color=(white) rgb255=(100,149,237)},
        xticklabels={Agensla,Androm,Convagent,Crypt,Crysan,DCRat,Injuke,Makoob,Mokes,Noon,Remcos,Seraph,SnakeLogger,Stealerc,Strab,Taskun,Zenpak},
        xtick={0,...,16}, xtick style={draw=none},
        xticklabel style={scale=0.85,anchor=east,rotate=60,yshift=-5pt,font=\tt},
        yticklabels={Agensla,Androm,Convagent,Crypt,Crysan,DCRat,Injuke,Makoob,Mokes,Noon,Remcos,Seraph,SnakeLogger,Stealerc,Strab,Taskun,Zenpak},
        ytick={0,...,16}, ytick style={draw=none},
        enlargelimits=false,
        yticklabel style={scale=0.85,font=\tt},
        colorbar, colorbar style={
            colorbar/width=2mm,
            ytick={0,30,60,90,120,150}, yticklabels={0,30,60,90,120,150},
            yticklabel={\pgfmathprintnumber\tick},
            yticklabel style={scale=0.9,/pgf/number format/fixed,/pgf/number format/fixed zerofill,/pgf/number format/precision=0}},
        colorbar=false,
        point meta min=0, point meta max=150,
        nodes near coords={\pgfmathprintnumber\pgfplotspointmeta},
        nodes near coords black white/.style={
            small value/.style={yshift=-5pt,text=black,/pgf/number format/fixed,/pgf/number format/precision=0,/pgf/number format/zerofill=true,scale=0.6},
            large value/.style={yshift=-5pt,text=white,/pgf/number format/fixed,/pgf/number format/precision=0,/pgf/number format/zerofill=true,scale=0.6},
            every node near coord/.style={check for zero/.code={
                \pgfmathfloatifflags{\pgfplotspointmeta}{0}{\pgfkeys{/tikz/coordinate}}{
                    \begingroup\pgfkeys{/pgf/fpu}\pgfmathparse{\pgfplotspointmeta<#1}
                    \global\let\result=\pgfmathresult\endgroup
                    \pgfmathfloatcreate{1}{1.0}{0}\let\ONE=\pgfmathresult
                    \ifx\result\ONE\pgfkeysalso{/pgfplots/small value}\else\pgfkeysalso{/pgfplots/large value}\fi}},check for zero}},
        nodes near coords black white=75,
    ]
\addplot[matrix plot, mesh/cols=17, point meta=explicit, draw=gray]
table [x=x, y=y, meta=C] {
x y C
0 0 11
1 0 0
2 0 0
3 0 6
4 0 1
5 0 0
6 0 0
7 0 0
8 0 1
9 0 1
10 0 0
11 0 0
12 0 42
13 0 2
14 0 0
15 0 86
16 0 0
0 1 5
1 1 12
2 1 0
3 1 6
4 1 2
5 1 0
6 1 1
7 1 14
8 1 6
9 1 1
10 1 1
11 1 4
12 1 48
13 1 4
14 1 0
15 1 45
16 1 1
0 2 2
1 2 3
2 2 59
3 2 3
4 2 0
5 2 0
6 2 29
7 2 0
8 2 6
9 2 7
10 2 0
11 2 15
12 2 6
13 2 3
14 2 0
15 2 9
16 2 8
0 3 7
1 3 1
2 3 0
3 3 15
4 3 0
5 3 2
6 3 4
7 3 0
8 3 1
9 3 1
10 3 0
11 3 4
12 3 23
13 3 0
14 3 0
15 3 92
16 3 0
0 4 5
1 4 1
2 4 2
3 4 6
4 4 85
5 4 0
6 4 4
7 4 0
8 4 0
9 4 2
10 4 0
11 4 4
12 4 5
13 4 2
14 4 0
15 4 34
16 4 0
0 5 11
1 5 0
2 5 0
3 5 20
4 5 0
5 5 101
6 5 0
7 5 0
8 5 0
9 5 0
10 5 0
11 5 0
12 5 1
13 5 0
14 5 0
15 5 17
16 5 0
0 6 12
1 6 3
2 6 2
3 6 6
4 6 2
5 6 2
6 6 20
7 6 0
8 6 2
9 6 6
10 6 4
11 6 37
12 6 7
13 6 7
14 6 0
15 6 40
16 6 0
0 7 0
1 7 2
2 7 0
3 7 0
4 7 0
5 7 0
6 7 0
7 7 124
8 7 0
9 7 0
10 7 0
11 7 1
12 7 18
13 7 3
14 7 0
15 7 2
16 7 0
0 8 1
1 8 0
2 8 1
3 8 2
4 8 0
5 8 6
6 8 24
7 8 0
8 8 71
9 8 0
10 8 1
11 8 0
12 8 6
13 8 12
14 8 0
15 8 3
16 8 23
0 9 11
1 9 1
2 9 0
3 9 4
4 9 0
5 9 0
6 9 1
7 9 0
8 9 0
9 9 9
10 9 4
11 9 3
12 9 45
13 9 0
14 9 0
15 9 72
16 9 0
0 10 1
1 10 4
2 10 0
3 10 7
4 10 1
5 10 1
6 10 2
7 10 2
8 10 0
9 10 5
10 10 44
11 10 3
12 10 28
13 10 6
14 10 0
15 10 46
16 10 0
0 11 8
1 11 0
2 11 0
3 11 6
4 11 0
5 11 0
6 11 4
7 11 0
8 11 0
9 11 2
10 11 0
11 11 22
12 11 19
13 11 0
14 11 1
15 11 88
16 11 0
0 12 2
1 12 0
2 12 0
3 12 0
4 12 0
5 12 0
6 12 0
7 12 0
8 12 0
9 12 0
10 12 0
11 12 2
12 12 108
13 12 0
14 12 0
15 12 38
16 12 0
0 13 7
1 13 7
2 13 3
3 13 12
4 13 1
5 13 8
6 13 9
7 13 2
8 13 14
9 13 5
10 13 1
11 13 2
12 13 8
13 13 43
14 13 0
15 13 22
16 13 6
0 14 14
1 14 18
2 14 0
3 14 8
4 14 0
5 14 8
6 14 2
7 14 5
8 14 6
9 14 0
10 14 2
11 14 2
12 14 53
13 14 11
14 14 10
15 14 10
16 14 1
0 15 10
1 15 0
2 15 0
3 15 3
4 15 1
5 15 0
6 15 0
7 15 0
8 15 0
9 15 0
10 15 0
11 15 0
12 15 43
13 15 0
14 15 0
15 15 93
16 15 0
0 16 5
1 16 2
2 16 2
3 16 2
4 16 0
5 16 2
6 16 8
7 16 1
8 16 41
9 16 4
10 16 0
11 16 1
12 16 7
13 16 14
14 16 0
15 16 1
16 16 60
};
\end{axis}
\end{tikzpicture}%
    }
    &
    \scalebox{0.9}{%
        % byteclass_hcurve — Grad-CAM Overlay Images (acc=0.400)
\begin{tikzpicture}[scale=1.0]
    \begin{axis}[
        width=7.0cm, height=7.0cm,
        colormap={bluewhite}{color=(white) rgb255=(100,149,237)},
        xticklabels={Agensla,Androm,Convagent,Crypt,Crysan,DCRat,Injuke,Makoob,Mokes,Noon,Remcos,Seraph,SnakeLogger,Stealerc,Strab,Taskun,Zenpak},
        xtick={0,...,16}, xtick style={draw=none},
        xticklabel style={scale=0.85,anchor=east,rotate=60,yshift=-5pt,font=\tt},
        yticklabels={Agensla,Androm,Convagent,Crypt,Crysan,DCRat,Injuke,Makoob,Mokes,Noon,Remcos,Seraph,SnakeLogger,Stealerc,Strab,Taskun,Zenpak},
        ytick={0,...,16}, ytick style={draw=none},
        enlargelimits=false,
        yticklabel style={scale=0.85,font=\tt},
        colorbar, colorbar style={
            colorbar/width=2mm,
            ytick={0,30,60,90,120,150}, yticklabels={0,30,60,90,120,150},
            yticklabel={\pgfmathprintnumber\tick},
            yticklabel style={scale=0.825,/pgf/number format/fixed,/pgf/number format/fixed zerofill,/pgf/number format/precision=0}},
        point meta min=0, point meta max=150,
        nodes near coords={\pgfmathprintnumber\pgfplotspointmeta},
        nodes near coords black white/.style={
            small value/.style={yshift=-5pt,text=black,/pgf/number format/fixed,/pgf/number format/precision=0,/pgf/number format/zerofill=true,scale=0.6},
            large value/.style={yshift=-5pt,text=white,/pgf/number format/fixed,/pgf/number format/precision=0,/pgf/number format/zerofill=true,scale=0.6},
            every node near coord/.style={check for zero/.code={
                \pgfmathfloatifflags{\pgfplotspointmeta}{0}{\pgfkeys{/tikz/coordinate}}{
                    \begingroup\pgfkeys{/pgf/fpu}\pgfmathparse{\pgfplotspointmeta<#1}
                    \global\let\result=\pgfmathresult\endgroup
                    \pgfmathfloatcreate{1}{1.0}{0}\let\ONE=\pgfmathresult
                    \ifx\result\ONE\pgfkeysalso{/pgfplots/small value}\else\pgfkeysalso{/pgfplots/large value}\fi}},check for zero}},
        nodes near coords black white=75,
    ]
\addplot[matrix plot, mesh/cols=17, point meta=explicit, draw=gray]
table [x=x, y=y, meta=C] {
x y C
0 0 24
1 0 4
2 0 0
3 0 5
4 0 1
5 0 1
6 0 0
7 0 1
8 0 0
9 0 4
10 0 0
11 0 3
12 0 13
13 0 0
14 0 0
15 0 94
16 0 0
0 1 22
1 1 7
2 1 0
3 1 9
4 1 3
5 1 0
6 1 0
7 1 19
8 1 8
9 1 0
10 1 6
11 1 6
12 1 10
13 1 1
14 1 1
15 1 56
16 1 2
0 2 1
1 2 1
2 2 91
3 2 6
4 2 2
5 2 1
6 2 14
7 2 1
8 2 6
9 2 2
10 2 4
11 2 1
12 2 1
13 2 4
14 2 1
15 2 2
16 2 12
0 3 21
1 3 1
2 3 1
3 3 24
4 3 1
5 3 6
6 3 0
7 3 1
8 3 0
9 3 0
10 3 0
11 3 8
12 3 11
13 3 0
14 3 0
15 3 76
16 3 0
0 4 4
1 4 0
2 4 0
3 4 7
4 4 89
5 4 0
6 4 1
7 4 0
8 4 0
9 4 1
10 4 2
11 4 3
12 4 8
13 4 3
14 4 2
15 4 0
16 4 30
0 5 0
1 5 0
2 5 0
3 5 3
4 5 1
5 5 90
6 5 0
7 5 0
8 5 0
9 5 0
10 5 0
11 5 51
12 5 0
13 5 0
14 5 2
15 5 0
16 5 3
0 6 7
1 6 2
2 6 11
3 6 4
4 6 5
5 6 0
6 6 30
7 6 3
8 6 9
9 6 6
10 6 9
11 6 12
12 6 6
13 6 9
14 6 0
15 6 34
16 6 3
0 7 0
1 7 0
2 7 0
3 7 1
4 7 0
5 7 0
6 7 0
7 7 147
8 7 0
9 7 0
10 7 0
11 7 0
12 7 1
13 7 0
14 7 0
15 7 0
16 7 1
0 8 1
1 8 0
2 8 0
3 8 0
4 8 0
5 8 0
6 8 0
7 8 0
8 8 100
9 8 0
10 8 8
11 8 0
12 8 1
13 8 5
14 8 0
15 8 2
16 8 33
0 9 20
1 9 2
2 9 0
3 9 8
4 9 1
5 9 0
6 9 1
7 9 3
8 9 0
9 9 9
10 9 5
11 9 3
12 9 11
13 9 0
14 9 0
15 9 86
16 9 1
0 10 13
1 10 1
2 10 0
3 10 5
4 10 1
5 10 0
6 10 2
7 10 14
8 10 1
9 10 4
10 10 45
11 10 5
12 10 7
13 10 4
14 10 1
15 10 47
16 10 0
0 11 8
1 11 4
2 11 2
3 11 8
4 11 0
5 11 0
6 11 4
7 11 1
8 11 0
9 11 5
10 11 2
11 11 39
12 11 5
13 11 1
14 11 0
15 11 70
16 11 1
0 12 16
1 12 1
2 12 0
3 12 4
4 12 0
5 12 0
6 12 0
7 12 1
8 12 1
9 12 2
10 12 0
11 12 1
12 12 70
13 12 0
14 12 2
15 12 50
16 12 2
0 13 1
1 13 2
2 13 3
3 13 10
4 13 0
5 13 5
6 13 1
7 13 3
8 13 25
9 13 0
10 13 17
11 13 3
12 13 3
13 13 46
14 13 4
15 13 14
16 13 13
0 14 0
1 14 0
2 14 2
3 14 2
4 14 0
5 14 3
6 14 0
7 14 31
8 14 2
9 14 1
10 14 39
11 14 4
12 14 2
13 14 21
14 14 35
15 14 0
16 14 8
0 15 27
1 15 0
2 15 0
3 15 4
4 15 0
5 15 0
6 15 0
7 15 0
8 15 0
9 15 2
10 15 1
11 15 3
12 15 16
13 15 0
14 15 0
15 15 97
16 15 0
0 16 1
1 16 0
2 16 1
3 16 3
4 16 1
5 16 0
6 16 0
7 16 2
8 16 45
9 16 1
10 16 7
11 16 1
12 16 1
13 16 7
14 16 1
15 16 1
16 16 78
};
\end{axis}
\end{tikzpicture}%
    }
    \\
    \adjustbox{scale=0.85}{(c) \byteclassHilbert\ (original images)}
    &
    \adjustbox{scale=0.85}{(d) \byteclassHilbert\ (overlay images)}
    \end{tabular}
    \caption{Confusion matrices}\label{fig:conf}
\end{figure}

\subsection{Progressive Feature Fusion Results}
% changed the title to this
% \subsection{Random Forest Results}

Motivated by the strong performance of the hybrid CNN-HOG-XGBoost pipeline,
the following three experiments are conducted with progressively richer feature representations,
using the CNN's penultimate layer as a feature vector for downstream classifiers.
\begin{itemize}
\item For each transformation, the trained overlay CNN is used as a feature extractor. 
The~256-dimensional output of the penultimate dense layer is extracted for every overlay image, 
and three classifiers are trained on these embeddings, namely, SVM (RBF kernel with~$C = 10$), 
Random Forest (500 trees), and XGBoost (500 estimators, learning rate~$0.05$). Feature scaling 
via \texttt{StandardScaler} is applied before SVM training.
\item For each transformation type and each sample, 
256-dim embeddings are extracted from both the original image model 
and the overlay model. 
These vectors are concatenated into a single 512-dim feature vector per sample, 
and three classifiers (SVM, Random Forest, and XGBoost) are trained.
\item Embeddings are extracted from all~16 trained models---eight 
transformations and two image types (original and overlay) per sample. 
For each malware sample, all~16 embeddings of size~256 are concatenated into a
single feature vector of length~4096. 
Only samples present across all~16 models are included, 
yielding 16,997 samples.\footnote{Three samples were excluded due to their Grad-CAM heatmaps 
being degenerate. The gradient of the predicted class score with respect to the final feature map was effectively 
zero for these samples, producing a flat heatmap with no spatial variation.} 
We train a Random Forest with~500 trees on these feature vectors.
\end{itemize}

Table~\ref{tab:rfperform} gives the best classifier test accuracy obtained for each image transformation type 
for both the~256-dimensional and 512-dimensional experiments. We observe that
Random Forest consistently outperforms SVM and XGBoost across all transformations 
in both settings. Furthermore, the~512-dimensional combined features improve 
over the~256-dimensional overlay-only features in every transformation, confirming 
that the original image model captures complementary information to the overlay model.

\begin{table}[!htb]
\centering
\caption{Test accuracy for 256-dim and 512-dim feature experiments}
\label{tab:rfperform}
\begin{adjustbox}{scale=0.775}
\begin{tabular}{lccccccc}
\toprule
\multirow{2}{*}{\raisebox{-2pt}{\textbf{Transformation}}} & \multirow{2}{*}{\raisebox{-2pt}{\textbf{CNN Overlay}}}
& \multicolumn{3}{c}{\textbf{256-dim (Overlay Only)}} 
& \multicolumn{3}{c}{\textbf{512-dim (Orig $\boldsymbol{\|}$ Overlay)}} \\
\cmidrule(lr){3-5} \cmidrule(lr){6-8}
%\textbf{Transformation} & \textbf{CNN Overlay} 
& & \textbf{SVM} & \ \ \ \ \textbf{RF} & \textbf{XGB}
& \textbf{SVM} & \ \ \ \ \textbf{RF} & \textbf{XGB} \\
\midrule
\grayscale          & 0.602 & 0.658 & \ \ \ \ 0.668 & 0.652 & 0.695 & \ \ \ \ \textbf{0.707} & 0.696 \\
\entropyHilbert    & 0.624 & 0.650 & \ \ \ \ 0.658 & 0.646 & 0.648 & \ \ \ \ \textbf{0.715} & 0.691 \\
\byteclassHilbert  & 0.400 & 0.570 & \ \ \ \ 0.638 & 0.628 & 0.503 & \ \ \ \ \textbf{0.653} & 0.639 \\
\byteclass          & 0.572 & 0.651 & \ \ \ \ 0.679 & 0.666 & 0.571 & \ \ \ \ \textbf{0.700} & 0.675 \\
\hit                & 0.548 & 0.622 & \ \ \ \ 0.651 & 0.644 & 0.581 & \ \ \ \ \textbf{0.686} & 0.677 \\
\spiral             & 0.499 & 0.518 & \ \ \ \ 0.546 & 0.533 & 0.482 & \ \ \ \ \textbf{0.568} & 0.559 \\
\bigramCartesian  & 0.566 & 0.569 & \ \ \ \ 0.576 & 0.568 & 0.564 & \ \ \ \ \textbf{0.617} & 0.613 \\
\bigramPolar      & 0.538 & 0.542 & \ \ \ \ 0.563 & 0.546 & 0.557 & \ \ \ \ \textbf{0.606} & 0.598 \\
\bottomrule
\end{tabular}
\end{adjustbox}
\end{table}

%\subsection{4096-dimensional Feature Results}

Table~\ref{tab:megafeatures} gives the result of training a Random Forest classifier 
on the full~4096-dimensional feature vector, which combines all~16 model embeddings (original
and overlay images for all eight image transformation types). An accuracy
of~0.777 is attained by this model, which provides a measurable improvement over all previous 
approaches considered in this chapter. We note that this result exceeds the best reported 
accuracy from prior work~\cite{stamp2024malwareimages}, which achieved~0.751
accuracy using HOG features with XGBoost, based on \grayscale\ image representations.

\begin{table}[!htb]
\centering
\caption{Random Forest results}
\label{tab:megafeatures}
\begin{adjustbox}{scale=0.85}
\begin{tabular}{l c}
\toprule
\textbf{Features} & \textbf{Test accuracy} \\
\midrule
4096-dimensional (all 16 models)  & \textbf{0.777} \\
512-dimensional best (\entropyHilbert) & 0.715 \\
HOG-XGBoost best (\hit) & 0.731 \\
\bottomrule
\end{tabular}
\end{adjustbox}
\end{table}

\begin{figure}[!htb]
    \centering
%    \resizebox{0.85\textwidth}{!}{%
    \begin{tikzpicture}[scale=1.0]
    \begin{axis}[
        width=10cm, height=10cm,
        colormap={bluewhite}{color=(white) rgb255=(100,149,237)},
        xticklabels={Agensla,Androm,Convagent,Crypt,Crysan,DCRat,Injuke,Makoob,Mokes,Noon,Remcos,Seraph,SnakeLogger,Stealerc,Strab,Taskun,Zenpak},
        xtick={0,...,16}, xtick style={draw=none},
        xticklabel style={scale=0.9,anchor=east,rotate=60,yshift=-5pt,font=\tt},
        yticklabels={Agensla,Androm,Convagent,Crypt,Crysan,DCRat,Injuke,Makoob,Mokes,Noon,Remcos,Seraph,SnakeLogger,Stealerc,Strab,Taskun,Zenpak},
        ytick={0,...,16}, ytick style={draw=none},
        enlargelimits=false,
        yticklabel style={scale=0.9,font=\tt},
        colorbar, colorbar style={
            colorbar/width=3.5mm,
            ytick={0,30,60,90,120,150}, yticklabels={0,30,60,90,120,150},
            yticklabel={\pgfmathprintnumber\tick},
            yticklabel style={scale=0.85,/pgf/number format/fixed,/pgf/number format/fixed zerofill,/pgf/number format/precision=0}},
        point meta min=0, point meta max=150,
        nodes near coords={\pgfmathprintnumber\pgfplotspointmeta},
        nodes near coords black white/.style={
            small value/.style={yshift=-6pt,text=black,/pgf/number format/fixed,/pgf/number format/precision=0,/pgf/number format/zerofill=true,scale=0.75},
            large value/.style={yshift=-6pt,text=white,/pgf/number format/fixed,/pgf/number format/precision=0,/pgf/number format/zerofill=true,scale=0.75},
            every node near coord/.style={check for zero/.code={
                \pgfmathfloatifflags{\pgfplotspointmeta}{0}{\pgfkeys{/tikz/coordinate}}{
                    \begingroup\pgfkeys{/pgf/fpu}\pgfmathparse{\pgfplotspointmeta<#1}
                    \global\let\result=\pgfmathresult\endgroup
                    \pgfmathfloatcreate{1}{1.0}{0}\let\ONE=\pgfmathresult
                    \ifx\result\ONE\pgfkeysalso{/pgfplots/small value}\else\pgfkeysalso{/pgfplots/large value}\fi}},check for zero}},
        nodes near coords black white=75,
    ]
\addplot[
  matrix plot,
  mesh/cols=17,
  point meta=explicit,
  draw=gray
] table [x=x, y=y, meta=C] {
x y C
0 0 91
1 0 1
2 0 0
3 0 8
4 0 0
5 0 0
6 0 4
7 0 0
8 0 0
9 0 15
10 0 0
11 0 7
12 0 2
13 0 2
14 0 0
15 0 20
16 0 0
0 1 3
1 1 86
2 1 1
3 1 2
4 1 1
5 1 0
6 1 7
7 1 13
8 1 4
9 1 3
10 1 1
11 1 15
12 1 1
13 1 6
14 1 3
15 1 3
16 1 1
0 2 1
1 2 1
2 2 120
3 2 0
4 2 3
5 2 0
6 2 4
7 2 0
8 2 4
9 2 0
10 2 0
11 2 1
12 2 1
13 2 1
14 2 7
15 2 0
16 2 7
0 3 2
1 3 1
2 3 1
3 3 115
4 3 0
5 3 0
6 3 1
7 3 0
8 3 0
9 3 6
10 3 0
11 3 13
12 3 0
13 3 6
14 3 0
15 3 5
16 3 0
0 4 0
1 4 0
2 4 0
3 4 5
4 4 122
5 4 0
6 4 4
7 4 0
8 4 0
9 4 0
10 4 0
11 4 17
12 4 0
13 4 1
14 4 1
15 4 0
16 4 0
0 5 0
1 5 0
2 5 1
3 5 0
4 5 0
5 5 144
6 5 1
7 5 0
8 5 0
9 5 0
10 5 0
11 5 0
12 5 0
13 5 1
14 5 2
15 5 0
16 5 0
0 6 0
1 6 0
2 6 0
3 6 4
4 6 2
5 6 0
6 6 102
7 6 0
8 6 0
9 6 4
10 6 2
11 6 17
12 6 1
13 6 14
14 6 2
15 6 0
16 6 2
0 7 0
1 7 0
2 7 0
3 7 0
4 7 0
5 7 0
6 7 0
7 7 149
8 7 0
9 7 0
10 7 0
11 7 0
12 7 0
13 7 0
14 7 1
15 7 0
16 7 0
0 8 1
1 8 0
2 8 0
3 8 0
4 8 0
5 8 1
6 8 0
7 8 0
8 8 120
9 8 0
10 8 0
11 8 0
12 8 0
13 8 6
14 8 2
15 8 0
16 8 20
0 9 16
1 9 1
2 9 0
3 9 7
4 9 0
5 9 0
6 9 8
7 9 0
8 9 0
9 9 84
10 9 5
11 9 8
12 9 2
13 9 0
14 9 1
15 9 18
16 9 0
0 10 5
1 10 1
2 10 0
3 10 4
4 10 2
5 10 0
6 10 4
7 10 3
8 10 0
9 10 6
10 10 101
11 10 13
12 10 0
13 10 0
14 10 4
15 10 7
16 10 0
0 11 2
1 11 0
2 11 0
3 11 3
4 11 2
5 11 0
6 11 4
7 11 0
8 11 0
9 11 2
10 11 0
11 11 136
12 11 0
13 11 1
14 11 0
15 11 0
16 11 0
0 12 5
1 12 0
2 12 0
3 12 2
4 12 0
5 12 0
6 12 3
7 12 0
8 12 0
9 12 0
10 12 0
11 12 4
12 12 129
13 12 1
14 12 3
15 12 3
16 12 0
0 13 0
1 13 1
2 13 2
3 13 6
4 13 1
5 13 0
6 13 7
7 13 2
8 13 9
9 13 0
10 13 1
11 13 4
12 13 0
13 13 113
14 13 2
15 13 0
16 13 2
0 14 0
1 14 0
2 14 0
3 14 0
4 14 0
5 14 0
6 14 1
7 14 5
8 14 0
9 14 0
10 14 1
11 14 0
12 14 0
13 14 8
14 14 134
15 14 0
16 14 1
0 15 11
1 15 0
2 15 0
3 15 5
4 15 0
5 15 0
6 15 1
7 15 0
8 15 0
9 15 10
10 15 1
11 15 2
12 15 0
13 15 0
14 15 0
15 15 120
16 15 0
0 16 0
1 16 1
2 16 0
3 16 0
4 16 0
5 16 0
6 16 0
7 16 0
8 16 15
9 16 0
10 16 0
11 16 1
12 16 0
13 16 14
14 16 4
15 16 0
16 16 115
};
\end{axis}
%\draw[black,thick] (2.625,5.8) circle(0.5);
%\draw[red,dashed,thick] (3.675,5.8) circle(0.5);
\end{tikzpicture}
%    }
\caption{Confusion matrix for 4096-dimensional features ($\mbox{accuracy} = 0.777$)}
\label{fig:rf_best4096}
\end{figure}
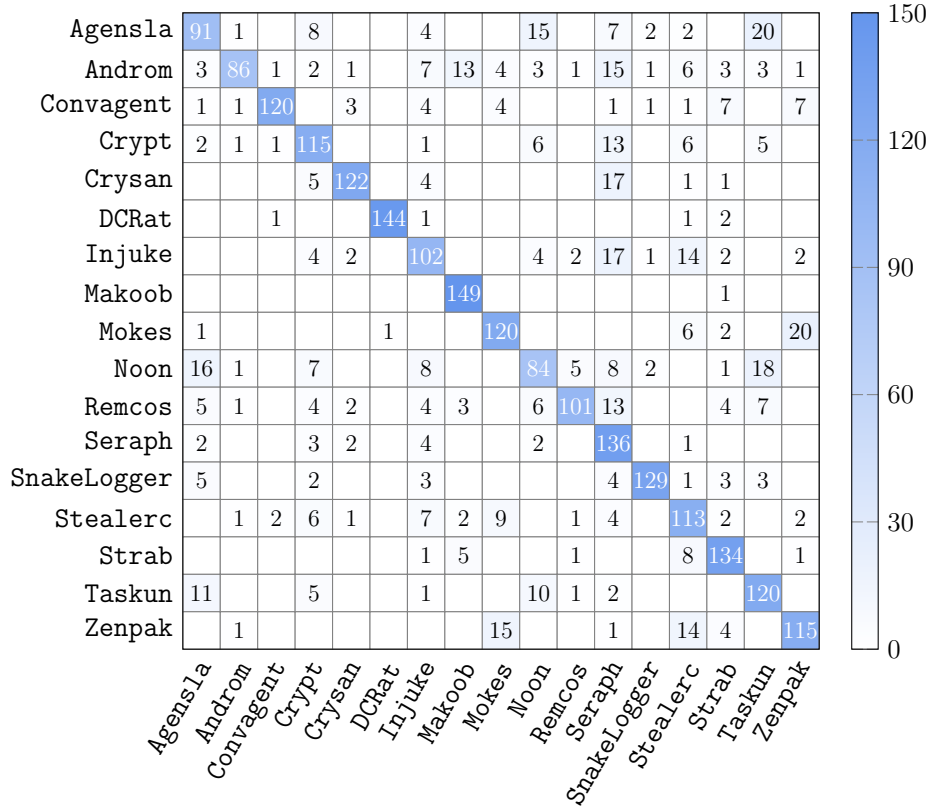

Of the~16,997 samples included (those present across all~8 transformations in both original and overlay form), 
the confusion matrix for the Random Forest trained on the~4096-dimensional combined embeddings 
%achieves a test accuracy of~0.777, as seen
is given in Figure~\ref{fig:rf_best4096}. As compared to the confusion
matrices given above, many improvements are readily apparent. 
For example, the malware families \texttt{Agensla} and \texttt{Taskun} are confused
with each other far less often in Figure~\ref{fig:rf_best4096} than in 
any of the previous confusion matrices.

%\begin{figure}[!htb]
%    \centering
%    \includegraphics[width=0.85\linewidth]{images/cm_mega_features_RF.png}
%    \caption{Confusion matrix for 4096-dimensional features ($\mbox{accuracy} = 0.777$)}
%    \label{fig:rf_best4096}
%\end{figure}

Per-family analysis of the confusion matrix in Figure~\ref{fig:rf_best4096}
reveals that \texttt{Makoob} is the most reliably classified family~(0.993), while \texttt{Noon}~(0.560) 
and \texttt{Androm}~(0.573) remain the most difficult, despite the richer feature representation. 
Table~\ref{tab:megafamily} gives the per-family accuracy for the five best and five 
worst performing families.
As in previous experiments, \texttt{Noon}, \texttt{Androm}, and \texttt{Injuke} appear at the bottom of these 
per-family rankings.
%across all experiments suggests that these families share structural binary characteristics that 
%are difficult to distinguish through image-based representations regardless of the model complexity 
%or feature richness used.

\begin{table}[!htb]
\centering
\caption{Best and worst accuracies for 4096-dimensional model} %(Best and Worst 5 Families)}
\label{tab:megafamily}
\begin{adjustbox}{scale=0.85}
\begin{tabular}{l c c c}
\toprule
\textbf{Family} & \textbf{Correct} & \textbf{Total} & \textbf{Accuracy} \\
\midrule
\texttt{Makoob}      & 149 & 150 & 0.993 \\
\texttt{DCRat}       & 144 & 150 & 0.960 \\
\texttt{Seraph}      & 136 & 150 & 0.907 \\
\texttt{Strab}       & 134 & 150 & 0.893 \\
\texttt{SnakeLogger} & 129 & 150 & 0.860 \\ \midrule
%\ \ \ $\vdots$ & $\vdots$ & $\vdots$ & $\vdots$ \\
\texttt{Zenpak}      & 115 & 150 & 0.767 \\
\texttt{Stealerc}    & 113 & 150 & 0.753 \\
\texttt{Injuke}      & 102 & 150 & 0.680 \\
\texttt{Androm}      & \zz86  & 150 & 0.573 \\
\texttt{Noon}        & \zz84  & 150 & 0.560 \\
\bottomrule
\end{tabular}
\end{adjustbox}
\end{table}

%\begin{table}[!htb]
%\centering
%\caption{Per-family accuracy (4096-dimensional Random Forest)} %(Best and Worst 5 Families)}
%\label{tab:megafamily}
%\begin{adjustbox}{scale=0.85}
%\begin{tabular}{l c c c}
%\toprule
%\textbf{Family} & \textbf{Correct} & \textbf{Total} & \textbf{Accuracy} \\
%\midrule
%\texttt{Agensla}
%\texttt{Androm}
%\texttt{Convagent}
%\texttt{Crypt}
%\texttt{Crysan}
%\texttt{DCRat}
%\texttt{Injuke}
%\texttt{Makoob}
%\texttt{Mokes}
%\texttt{Noon}
%\texttt{Remcos}
%\texttt{Seraph}
%\texttt{SnakeLogger}
%\texttt{Stealerc}
%\texttt{Strab}
%\texttt{Taskun}
%\texttt{Zenpak}
%\bottomrule
%\end{tabular}
%\end{adjustbox}
%\end{table}

%\subsection{Classification Accuracy Progression}

Table~\ref{tab:progression} summarizes the accuracy progression across our experiments.
As previously mentioned, in prior work~\cite{stamp2024malwareimages} involving this
same dataset, the best accuracy obtained was~0.751. We have extended this previous work,
achieving a consistent progression through various feature combinations, with a~4096-dimensional 
Random Forest combining all~16 trained models achieving an accuracy of~0.777, 
thereby exceeding the previous benchmark.

\begin{table}[!htb]
\centering
\caption{Classification accuracy across all experiments}
\label{tab:progression}
\begin{adjustbox}{scale=0.85}
\begin{tabular}{l c}
\toprule
\textbf{Model / Approach} & \textbf{Test accuracy} \\
\midrule
XGBoost-HOG (\grayscale)~\cite{stamp2024malwareimages} & 0.751 \\
%\midrule
CNN on original images 
	& 0.688 \\
CNN on Grad-CAM overlays %(CS297)
	& 0.604 \\
CNN-HOG-XGBoost on overlays %(CS297)
	& 0.731 \\
256-dim RF on overlay CNN features %(CS298)
	& 0.679 \\
512-dim RF: orig + overlay features %(CS298)
	& 0.715 \\
4096-dim RF: all 16 models %(CS298)
	& \textbf{0.777} \\
\bottomrule
\end{tabular}
\end{adjustbox}
\end{table}

\section{Discussion}\label{chap:discussion}

Across raw images and Grad-CAM overlays, \entropyHilbert\ and \grayscale\ consistently 
produce higher family-classification accuracy than \spiral, \byteclassHilbert, or \hit. 
These results indicate that
transformations that preserve spatial locality or encode entropy-related structure tend 
to yield clearer visual patterns that CNNs can exploit.
Entropy-based and grayscale-based representations also yield more concentrated and 
interpretable CAM regions, while \spiral\ and \byteclassHilbert\ often produce diffuse 
or noisy patterns that are less amenable to XAI-based analysis.

Random Forest and XGBoost outperform the Linear SVC on Grad-CAM feature vectors, 
demonstrating that geometric and texture-based descriptors (LBP, Hu moments, GLCM) capture useful information. 
These results highlight that classical feature engineering can still contribute valuable information, even 
when image-based analysis is employed.

For every image transformation type, concatenating CNN embeddings with HOG descriptors and training 
an XGBoost classifier yields substantially higher accuracy than a comparable CNN-only model. 
These gains are not limited to the strongest transformations (i.e., \entropyHilbert\ and \grayscale), but 
extend to weaker representations as well (e.g., \spiral, \byteclass, and \byteclassHilbert).

Notably, \hit\ achieves the highest hybrid test accuracy~(0.731), despite being only a mid-range 
performer in the CNN-only setting.  Similarly, \byteclass\ and \byteclassHilbert, which show
relatively weak CNN performance, are among the strongest transformations in the hybrid 
pipeline (accuracies of~0.721 and~0.718, respectively). These results indicate that HOG 
captures structural and gradient-based information that CNN embeddings alone do not fully 
exploit, particularly for transformations that emphasize texture or local transitions.

Overall, our hybrid approach confirms that deep and classical feature-learning methods 
provide complementary strengths. Specifically, CNN embeddings encode high-level semantics, 
while HOG captures fine-grained local patterns. XGBoost effectively integrates these 
heterogeneous descriptors, leading to consistent and often dramatic improvements across 
all image transformations. Furthermore, our results highlight the interplay among transformation choice, XAI, 
and classification accuracy---some transformations yield more interpretable and discriminative 
CAM patterns, while others yield greater accuracy.

The qualitative comparison between Grad-CAM and HiResCAM confirms that for MobileNetV2 
applied to malware imagery, the two methods produce effectively equivalent explanations at the 
final convolutional layer. As mentioned above, 
this may be attributable to the small spatial dimensions of MobileNetV2's 
final feature map ($7 \times 7$), which reduces the difference between global average pooling of 
gradients (Grad-CAM) and element-wise gradient-activation products (HiResCAM). The subtle 
differences observed in the difference maps suggest that HiResCAM provides marginally finer spatial 
resolution in the attention boundaries, but not sufficiently distinct to warrant a separate evaluation 
pipeline for this architecture and task.
%Future work using architectures with larger final feature maps, or applying the comparison 
%at intermediate layers, may reveal more meaningful differences between the two methods.

A central finding of the faithfulness evaluation is that classification accuracy and 
explanation faithfulness do not align. For example, \bigramPolar\ ranks seventh with respect to
CNN test accuracy, but first on faithfulness score. Conversely, \grayscale\ achieves good CNN accuracy, 
but ranks seventh on faithfulness. Among the eight image transformation types, only 
\entropyHilbert\ performs strongly on both---it has the highest CNN accuracy and 
the second-highest faithfulness score---making it the best single transformation for 
applications that require both accuracy and interpretability.
These results demonstrate that in the malware domain,
choosing an image transformation based solely on accuracy risks overlooking 
transformations that provide more interpretable and trustworthy indicators of model behavior.

Recall that faithfulness 
measures whether the explanation correctly identifies what the model uses to make a decision, 
while stability measures whether the explanation is reproducible under small input changes.
The inverse relationship between faithfulness and stability rankings suggests that these two 
properties of explanation quality capture distinct and complementary aspects.  A highly faithful 
explanation may pinpoint a precise and sensitive region of the image that is genuinely discriminative, 
but also easily perturbed, leading to lower stability.
For practical deployment in security analysis, both of these properties matter, as an analyst would 
want explanations that are both correct (i.e., faithful) and reliable (i.e., stable). Based on these
criteria, \entropyHilbert\ may offer the best practical compromise, while \spiral\ is the only 
transformation that ranks poorly with respect to both of these properties.

Our experiments further demonstrate that Grad-CAM overlay images are not merely 
a visualization tool but can serve as an enhanced image representation for classification. 
For five of the eight transformations, training a CNN on overlay images exceeds 
the performance of models trained only on the original images. This improvement is most 
pronounced for \grayscale\ and \byteclassHilbert.
%where the overlay adds approximately~5-to-6 percentage points of accuracy.
In contrast, for the bigram-based transformations, overlay images perform slightly worse 
than the original images. This is consistent with the observation that bigram representations 
encode frequency-domain structure in a way that may be partially obscured by the heatmap overlay. 
These transformations may benefit from alternative blending strategies, or from using the 
raw CAM heat maps rather than the overlay images.

The progressive improvement from~256-dimensional to~512-dimensional to~4096-dimensional 
features demonstrates that different models trained on different image representations capture 
complementary information. A key insight is that combining original image embeddings with 
overlay image embeddings (512-dimension) is consistently better than using either alone, 
again indicating that the overlay images provide additional context that is not available directly from
the original images.

The~4096-dimensional experiment extends this logic to its natural limit, combining 
all~16 available representations. As noted above, the resulting accuracy of~0.777 exceeds 
the previous benchmark for this same dataset. This suggests that feature combinations 
across multiple transformation types is a viable strategy for improved malware classification,
and potentially complementary to the handcrafted HOG features used in prior work.

\texttt{Noon}, \texttt{Androm}, \texttt{Agensla}, and \texttt{Injuke} appear in the bottom five families by accuracy across virtually 
every experiment. Even with all~16 models combined, \texttt{Noon} achieves only~0.560 accuracy 
and \texttt{Androm}~0.573. This consistency across approaches suggests that these families share 
characteristics that make them inherently difficult to distinguish 
using image-based features alone. 
%Future work could investigate whether these families benefit 
%from behavioral or static analysis features that are not captured in image representations.

The exclusive use of MobileNetV2 as a backbone is a 
limitation of our research. Another obvious limitation is 
the computational cost of generating full Grad-CAM datasets.
%and the absence of quantitative consistency metrics for explanations.

\section{Conclusion and Future Work}\label{chap:conclusion}
 
This chapter provides a comprehensive investigation of malware image transformations, 
their impact on convolutional neural network performance, and their structure revealed through 
Grad-CAM visualizations. Initially, we
established a foundation by generating large-scale Grad-CAM datasets across eight image 
transformations and seventeen malware families. We then trained per-transformation CNN classifiers 
on both original and overlay images, extracting handcrafted features from Grad-CAM heatmaps, and 
developing a hybrid CNN-HOG-XGBoost pipeline that achieved a test accuracy of~0.731. These 
results confirmed that transformation choice strongly influences classification performance and 
that Grad-CAM overlays preserve meaningful family-discriminative structure.
 
We then extended this foundation in four directions. First, HiResCAM was
compared against Grad-CAM, revealing that the two methods produce equivalent explanations 
at the final convolutional layer of MobileNetV2, perhaps due to the small spatial dimensions of that 
layer's feature map. Second, quantitative faithfulness and stability metrics were introduced and 
evaluated across all eight transformations. These results revealed that for malware images,
accuracy and interpretability do not align. 
  
A progressive feature combination strategy was developed. Extracting~256-dimensional 
CNN embeddings from overlay models and training Random Forest improved over the 
CNN baseline but fell short of the previous benchmark. Finally, combining all~16 model embeddings 
(original and overlay for each of the eight transformations) into a~4096-dim feature vector and training 
a single Random Forest classifier achieved a test accuracy that exceeded the previous benchmark.
 
Taken together, these findings contribute new insights into the relationship between malware image 
transformation choice, deep learning-based classification, and explanation quality. They 
demonstrate that combining complementary deep feature representations across multiple transformation 
types is a useful strategy for malware classification, that explanation faithfulness and stability are distinct 
and measurable properties that do not necessarily correlate with accuracy, and that Grad-CAM overlays 
can serve as both interpretability tools and as enhanced input representations. These contributions offer 
a foundation for building malware classification systems that are simultaneously accurate, 
interpretable, and explainable.
 
There are several promising directions for extending the work considered in this chapter, 
We showed that combining HOG descriptors with CNN embeddings consistently improved 
the accuracy over CNN-only models. A natural next step is to add HOG features on top of 
our~512-dimensional and 4096-dimensional feature vectors to determine whether handcrafted 
texture descriptors provide additional discriminative power in these cases.
  
Our comparison between Grad-CAM and HiResCAM at the final convolutional layer of MobileNetV2 
showed nearly equivalent explanations. It would be interesting to apply both methods at intermediate 
layers of the backbone, where feature maps are larger and the element-wise formulation of HiResCAM 
would be expected to produce more spatially precise explanations. This could reveal transformation-specific 
differences in explanation quality that are not visible at the final layer. In addition, other XAI techniques
could be considered, such as Grad-CAM\texttt{++}, Integrated Gradients, or SHAP applied directly to 
convolutional activations. Comparing these methods would help determine whether the faithfulness 
and stability patterns observed for Grad-CAM are consistent across explanation frameworks.
 
Understanding the robustness of XAI explanations in adversarial settings is critical 
for practical malware analysis. Open research problems in this area include evaluating whether 
adversarial perturbations can alter Grad-CAM heatmaps without changing predictions, studying 
whether attackers can intentionally mislead or obscure high-importance regions, and analyzing 
which transformations are most resilient to such adversarial manipulation.

%\begin{figure}[!htb]
%    \centering
%    \input figures/conf_example2.tex
%    \caption{Example confusion matrix}\label{fig:conf}
%\end{figure}

\bibliographystyle{plain}
\bibliography{references}

@book{aycock,
title={Computer Viruses and Malware},
year={2006},
publisher={Springer},
author={John Aycock}
}

@misc{virusshare,
title = {{VirusShare}},
key = {VirusShare},
year = {2026},
howpublished = {\url{https://www.virusshare.com/}}
}

@misc{bazaar,
title = {{MalwareBazaar}},
key = {MalwareBazaar},
year = {2026},
howpublished = {\url{https://bazaar.abuse.ch/}}
}

@misc{VXU,
title = {{VX Underground}},
key = {VX Underground},
year = {2026},
howpublished = {\url{https://vx-underground.org/}}
}

@inproceedings{Bhodia19,
  author       = {Niket Bhodia and
                  Pratikkumar Prajapati and
                  Fabio Di Troia and
                  Mark Stamp},
  editor       = {Paolo Mori and
                  Steven Furnell and
                  Olivier Camp},
  title        = {Transfer Learning for Image-based Malware Classification},
  booktitle    = {Proceedings of the 5th International Conference on Information Systems
                  Security and Privacy}, 
  series = {ICISSP}, 
  pages        = {719--726},
  year         = {2019},
}

@inproceedings{Yajamanam18,
  author       = {Sravani Yajamanam and
                  Vikash Raja Samuel Selvin and
                  Fabio Di Troia and
                  Mark Stamp},
  editor       = {Paolo Mori and
                  Steven Furnell and
                  Olivier Camp},
  title        = {Deep Learning versus {G}ist Descriptors for Image-based Malware Classification},
  booktitle    = {Proceedings of the 4th International Conference on Information Systems
                  Security and Privacy}, 
  series = {ICISSP}, 
  pages        = {553--561},
  year         = {2018},
}

@incollection{Stamp2021,
author={Stamp, Mark},
editor={Stamp, Mark and Alazab, Mamoun and Shalaginov, Andrii},
title={A Selective Survey of Deep Learning Techniques and Their Application to Malware Analysis},
booktitle={Malware Analysis Using Artificial Intelligence and Deep Learning},
year={2021},
publisher={Springer},
pages={3--51},
}

@article{wing,
  author       = {Wing Wong and
                  Mark Stamp},
  title        = {Hunting for metamorphic engines},
  journal      = {Journal in Computer Virology},
  volume       = {2},
  number       = {3},
  pages        = {211--229},
  year         = {2006},
  }

@misc{kasperskyandrom,
      title={{Androm: Backdoor.Win32.Androm}}, 
      key={Androm},
      year={2025},
      howpublished={\url{https://threats.kaspersky.com/en/threat/Backdoor.Win32.Androm/}}, 
}

@misc{kasperskycrypt,
      title={{Crypt: HackTool.Win32.Crypt.au}}, 
      key={Crypt},
      year={2025},
      howpublished={\url{https://threats.kaspersky.com/en/threat/HackTool.Win32.Crypt.au/}}, 
}

@misc{kasperskycrysan,
      title={{Crysan: Backdoor.Win32.Crysan.gen}}, 
      key={Crysan},
      year={2025},
      howpublished={\url{https://threats.kaspersky.com/en/threat/VHO:Backdoor.Win32.Crysan.gen/}}, 
}

@misc{kasperskydcrat,
      title={{DCRat: Backdoor.MSIL.DCRat.aae}}, 
      key={DCRat},
      year={2025},
      howpublished={\url{https://threats.kaspersky.com/en/threat/Backdoor.MSIL.DCRat.aae/}}, 
}

@misc{kasperskyinjuke,
      title={{Injuke: Trojan.Win32.Injuke.gen}}, 
      key={Injuke},
      year={2025},
      howpublished={\url{https://threats.kaspersky.com/en/threat/HEUR:Trojan.Win32.Injuke.gen/}}, 
}

@misc{kasperskymokes,
      title={{Mokes: Backdoor.Win32.Mokes}}, 
      key={Mokes},
      year={2025},
      howpublished={\url{https://threats.kaspersky.com/en/threat/Backdoor.Win32.Mokes/}}, 
}

@misc{kasperskynoon,
      title={{Noon: Trojan-Spy.Win32.Noon}}, 
      key={Noon},
      year={2025},
      howpublished={\url{https://threats.kaspersky.com/en/threat/Trojan-Spy.Win32.Noon/}}, 
}

@misc{kasperskyremcos,
      title={{Remcos: Backdoor.Win32.Remcos.aaaa}}, 
      key={Remcos},
      year={2025},
      howpublished={\url{https://threats.kaspersky.com/en/threat/Backdoor.Win32.Remcos.aaaa/}}, 
}

@misc{kasperskyseraph,
      title={{Seraph: Trojan-Downloader.Win32.Seraph.gen}}, 
      key={Seraph},
      year={2025},
      howpublished={\url{https://threats.kaspersky.com/en/threat/HEUR:Trojan-Downloader.Win32.Seraph.gen/}}, 
}

@misc{kasperskysnakelogger,
      title={{Snakelogger: Trojan-Spy.Win32.SnakeLogger.gen}}, 
      key={Snakelogger},
      year={2025},
      howpublished={\url{https://threats.kaspersky.com/en/threat/HEUR:Trojan-Spy.Win32.SnakeLogger.gen/}}, 
}

@misc{kasperskystealerc,
      title={{Stealerc: Trojan-PSW.Win32.Stealerc.gen}}, 
      key={Stealerc},
      year={2025},
      howpublished={\url{https://threats.kaspersky.com/en/threat/HEUR:Trojan-PSW.Win32.Stealerc.gen/}}, 
}

@misc{kasperskystrab,
      title={{Strab: Trojan.Win32.Strab.gen}}, 
      key={Strab},
      year={2025},
      howpublished={\url{https://threats.kaspersky.com/en/threat/HEUR:Trojan.Win32.Strab.gen/}}, 
}

@misc{kasperskytaskun,
      title={{Taskun: Trojan-Downloader.Win32.Taskun.gen}}, 
      key={Taskun},
      year={2025},
      howpublished={\url{https://threats.kaspersky.com/en/threat/HEUR:Trojan-Downloader.Win32.Taskun.gen/}}, 
}

@misc{kasperskyzenpak,
      title={{Zenpak: Trojan.Win32.Zenpak.gen}}, 
      key={Zenpak},
      year={2025},
      howpublished={\url{https://threats.kaspersky.com/en/threat/HEUR:Trojan.Win32.Zenpak.gen/}}, 
}

@article{CHEN2026131781,
title = {Dynamic malware detection based on enhanced semantic {API} sequence features},
journal = {Expert Systems with Applications},
volume = {315},
pages = {131781},
year = {2026},
author = {Zhiguo Chen and Lei Zhou and Qingcheng Liu and Weizhi Meng and Jian Weng},
}

@article{Manokaran03042023,
author = {Jadshan Manokaran and Gurusami Vairavel},
title = {{GIWRF-SMOTE}: Gini impurity-based weighted random forest with {SMOTE} 
for effective malware attack and anomaly detection in {IoT-Edge}},
journal = {Smart Science},
volume = {11},
number = {2},
pages = {276--292},
year = {2023}
}

@article{TANG2023103118,
title = {{BHMDC}: A byte and hex n-gram based malware detection and classification method},
journal = {Computers \&\ Security},
volume = {128},
pages = {103118},
year = {2023},
author = {Yonghe Tang and Xuyan Qi and Jing Jing and Chunling Liu and Weiyu Dong}
}

@article{MIMURA2022100521,
title = {Evaluation of printable character-based malicious {PE} file-detection method},
journal = {Internet of Things},
volume = {19},
pages = {100521},
year = {2022},
author = {Mamoru Mimura},
}

@article{BaysaLS13,
  author       = {Donabelle Baysa and
                  Richard M. Low and
                  Mark Stamp},
  title        = {Structural entropy and metamorphic malware},
  journal      = {Journal of Computer Virology and Hacking Techniques},
  volume       = {9},
  number       = {4},
  pages        = {179--192},
  year         = {2013},
}

@article{DamodaranS17,
  author       = {Anusha Damodaran and
                  Fabio Di Troia and
                  Corrado Aaron Visaggio and
                  Thomas H. Austin and
                  Mark Stamp},
  title        = {A comparison of static, dynamic, and hybrid analysis for malware detection},
  journal      = {Journal of Computer Virology and Hacking Techniques},
  volume       = {13},
  number       = {1},
  pages        = {1--12},
  year         = {2017}
}

@inproceedings{nataraj2011malware,
  title={Malware images: Visualization and automatic classification},
  author={Nataraj, Lakshmanan and Karthikeyan, Shankarapani and Jacob, George and Manjunath, BS},
  booktitle={Proceedings of the 8th International Symposium on Visualization for Cyber Security},
  pages={1--7},
  year={2011}
}

@inproceedings{entropy2022,
  author={Ling, Yeong Tyng and Phang, Piau and Chiew, Kang Leng and Zhang, Xiaowei},
  booktitle={2022 International Conference on Digital Transformation and Intelligence},
  series={ICDI}, 
  title={Malware Detection with Structural Entropy Features Using Multilayer Perceptron Neural Network}, 
  year={2022},
  pages={01--07},
}

@inproceedings{selvaraju2017gradcam,
  author={Selvaraju, Ramprasaath R. and Cogswell, Michael and Das, Abhishek and Vedantam, Ramakrishna and Parikh, Devi and Batra, Dhruv},
  booktitle={2017 IEEE International Conference on Computer Vision},
  series={ICCV}, 
  title={{Grad-CAM}: Visual Explanations from Deep Networks via Gradient-Based Localization}, 
  year={2017},
  pages={618--626},
}

@article{xai,
  author={Manthena, Harikha and Shajarian, Shaghayegh and Kimmell, Jeffrey C. and Abdelsalam, Mahmoud 
  	and Khorsandroo, Sajad and Gupta, Maanak},
  journal={IEEE Access}, 
  title={Explainable Artificial Intelligence ({XAI}) for Malware Analysis: A Survey of Techniques, Applications, and Open Challenges}, 
  year={2025},
  volume={13},
  pages={61611--61640},
}

@inproceedings{stamp2024malwareimages,
  author       = {Rishit Agrawal and
                  Kunal Bhatnagar and
                  Andrew Do and
                  Ronnit Rana and
                  Martin Jure\v{c}ek and
                  Mark Stamp},
  editor       = {Roberto Di Pietro and
                  Karen Renaud and
                  Paolo Mori},
  title        = {A Comparison of Selected Image Transformation Techniques for Malware Classification},
  booktitle    = {Proceedings of the 12th International Conference on Information Systems Security and Privacy}, 
  series = {ICISSP},
  volume = {2},
  pages        = {334--344},
  year         = {2026}
}

@article{grad_cam, 
title={Advancing malware imagery classification with explainable deep learning: 
	A state-of-the-art approach using {SHAP}, {LIME} and {Grad-CAM}},
  author={Sadia Nazim and Muhammad Mansoor Alam and Syed Safdar Rizvi 
  	and Jawahir Che Mustapha and Syed Shujaa Hussain and Mazliham Mohd Suud},
  journal={PLOS ONE},
  volume={20},
  number={5},
  pages={e0318542},
  year={2025}
}

@misc{rawmaltf,
      title={{RawMal-TF}: Raw Malware Dataset Labeled by Type and Family}, 
      author={David B\'{a}lik and Martin Jure\v{c}ek and Mark Stamp},
      year={2025},
      howpublished={\url{https://arxiv.org/abs/2506.23909}}, 
}

@inproceedings{hu_moment,
  author={Singh, Brajesh Kumar and Rai, Amrita and Kundu, Krishanu and Kalita, Karabi and Agrawal, Reshu},
  booktitle={2024 11th International Conference on Reliability, Infocom Technologies and Optimization 
  	(Trends and Future Directions)},
  series={ICRITO}, 
  title={An Empirical Analysis of Invariance {H}u's Moment Feature over a Digital Image}, 
  year={2024},
  pages={1--5},
}

@article{lbp,
  author={Timo Ojala and Matti Pietik\"{a}inen and Topi M\"{a}enp\"{a}\"{a}},
  journal={IEEE Transactions on Pattern Analysis and Machine Intelligence}, 
  title={Multiresolution gray-scale and rotation invariant texture classification with local binary patterns}, 
  year={2002},
  volume={24},
  number={7},
  pages={971--987},
}

@article{glcm,
  author={Haralick, Robert M. and Shanmugam, K. and Dinstein, Its'Hak},
  journal={IEEE Transactions on Systems, Man, and Cybernetics}, 
  title={Textural Features for Image Classification}, 
  year={1973},
  volume={SMC-3},
  number={6},
  pages={610--621},
}

@inproceedings{HOG,
  author={Navneet Dalal and Bill Triggs},
  booktitle={2005 IEEE Computer Society Conference on Computer Vision and Pattern Recognition},
  series={CVPR'05}, 
  title={Histograms of oriented gradients for human detection}, 
  year={2005},
  volume={1},
  pages={886--893},
}

@misc{hirescam2020,
  title={Use {HiResCAM} Instead of {Grad-CAM} for Faithful Explanations of Convolutional Neural Networks},
  author={Draelos, Rachel Lea and Carin, Lawrence},
  year={2020},
  howpublished={\url{https://arxiv.org/abs/2011.08891}}
}

@article{hama2023deletion,
  title={Deletion and Insertion Tests in Regression Models},
  author={Hama, Naofumi and Mase, Masayoshi and Owen, Art B.},
  journal={Journal of Machine Learning Research},
  volume={24},
  pages={1--38},
  year={2023},
}

@article{jei2023saliency,
author = {Tristan Gomez and Harold Mouch{\`e}re},
title = {Computing and evaluating saliency maps for image classification: a tutorial},
volume = {32},
journal = {Journal of Electronic Imaging},
number = {2},
pages = {020801},
year = {2023},
}

@inproceedings{petsiuk2018rise,
  title={{RISE}: Randomized Input Sampling for Explanation of Black-box Models},
  author={Petsiuk, Vitali and Das, Abir and Saenko, Kate},
  booktitle={British Machine Vision Conference},
  series={BMVC},
  pages={151},
  year={2018}
}

@inproceedings{chakraborty2022gradcam,
  title={Generalizing Adversarial Explanations with {Grad-CAM}},
  author={Chakraborty, Tanmay and others},
  booktitle={Proceedings of the IEEE/CVF Conference on Computer Vision and Pattern Recognition Workshops},
  series={CVPRW},
  pages={187--193},
  year={2022},
}

@misc{agensla_kaspersky,
  title = {{Agensla: Trojan-PSW.MSIL.Agensla}},
  key = {Agensla},
  year = {2025},
  howpublished = {\url{https://threats.kaspersky.com/en/threat/Trojan-PSW.MSIL.Agensla/}},
}

@misc{convagent,
  title = {{Convagent: Trojan.Win32.CONVAGENT.0NA103AQ24}},
  key = {Convagent},
  year = {2024},
  howpublished = {\url{https://www.trendmicro.com/vinfo/us/threat-encyclopedia/malware/Trojan.Win32.CONVAGENT.0NA103AQ24}},
}

@misc{varela2017bigram,
  author = {Mart\'{i}n Varela},
  title = {Simple Binary Data Visualization},
  year = {2017},
  howpublished = {\url{https://martin.varela.fi/2017/09/09/simple-binary-data-visualization/}},
}

@misc{chen2016xgboost,
  author       = {Tianqi Chen and
                  Carlos Guestrin},
  title        = {{XGBoost}: {A} Scalable Tree Boosting System},
  year         = {2016},
  howpublished         = {\url{http://arxiv.org/abs/1603.02754}},
}

@article{breiman2001random,
  title={Random Forests},
  author={Breiman, Leo},
  journal={Machine learning},
  volume={45},
  number={1},
  pages={5--32},
  year={2001},
  publisher={Springer},
  doi={10.1023/A:1010933404324}
}

%\section*{Appendix A}\label{app:a}
\section*{Appendix}\label{app:a}

\titleformat{\section}{\normalfont\large\bfseries}{}{0em}{#1\ \thesection}
\setcounter{section}{0}
\renewcommand{\thesection}{\Alph{section}}
\renewcommand{\thesubsection}{A.\arabic{subsection}}
\setcounter{table}{0}
\renewcommand{\thetable}{A.\arabic{table}}
\setcounter{figure}{0}
\renewcommand{\thefigure}{A.\arabic{figure}}

\begin{figure}[!htb]
    \centering
    \begin{tabular}{>{\centering\arraybackslash}m{0.2\linewidth} >{\centering\arraybackslash}m{0.2\linewidth} 
    	>{\centering\arraybackslash}m{0.2\linewidth} >{\centering\arraybackslash}m{0.2\linewidth}}
    \phantom{MMMMMMM} & \phantom{MMMMMMM} & \phantom{MMMMMMM} & \phantom{MMMMMMM} \\
    \multicolumn{4}{c}{\includegraphics[width=0.8\linewidth]{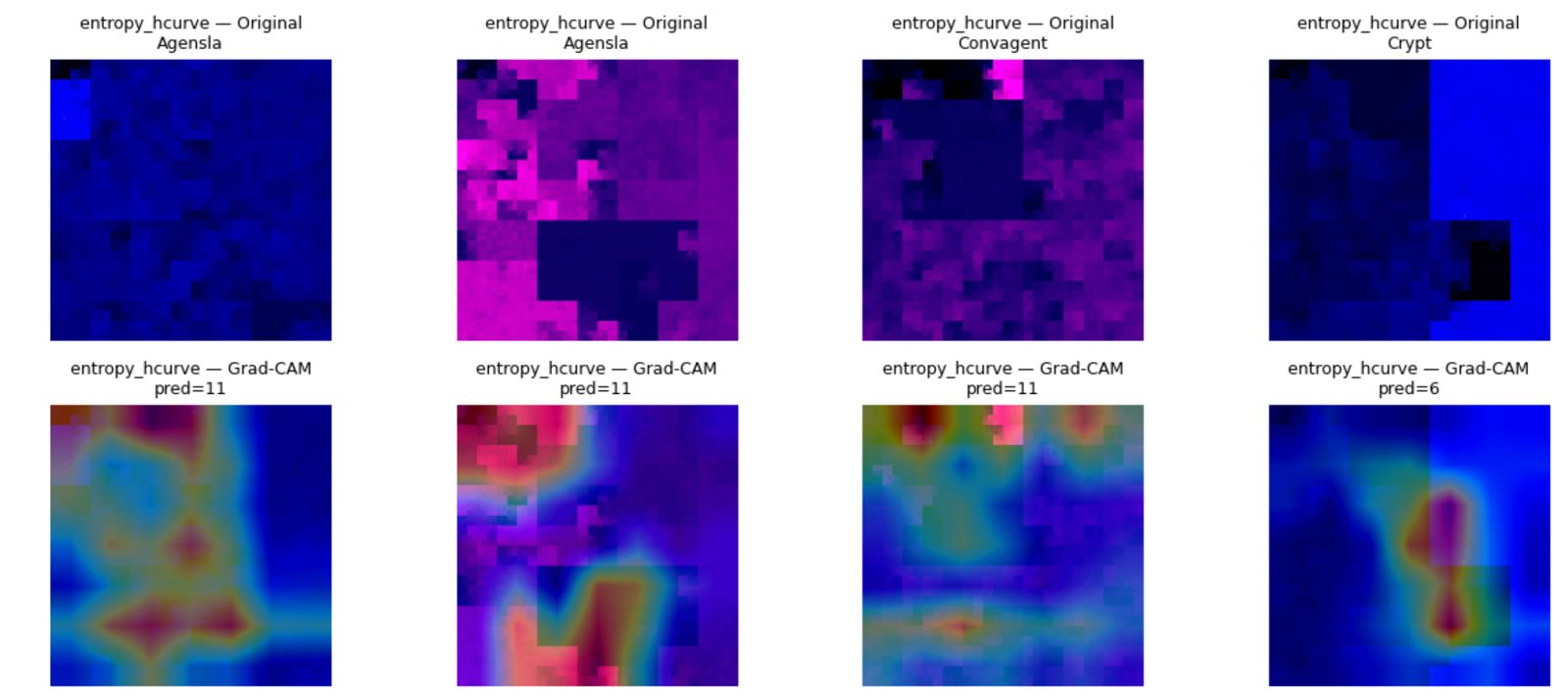}} \\[-1ex]
    \adjustbox{scale=0.85}{\ \ \ \ \texttt{Agensla}} & \adjustbox{scale=0.85}{\ \ \ \ \texttt{Agensla}} 
    	& \adjustbox{scale=0.85}{\ \ \texttt{Convagent}} & \adjustbox{scale=0.85}{\texttt{Crypt}} \\
%    \adjustbox{scale=1.0}{\ \ \ \ \ \ \ \ \texttt{Agensla}} & \adjustbox{scale=1.0}{\ \ \ \ \texttt{Agensla}} 
%    	& \adjustbox{scale=1.0}{\texttt{Convagent}} & \adjustbox{scale=1.0}{\!\!\!\!\!\!\!\!\texttt{Crypt}} \\
    \end{tabular}
    \caption{\entropyHilbert\ malware images and Grad-CAM overlays}\label{fig:gradcam2a}
\end{figure}

\begin{figure}[!htb]
    \centering
    \begin{tabular}{>{\centering\arraybackslash}m{0.2\linewidth} >{\centering\arraybackslash}m{0.2\linewidth} 
    	>{\centering\arraybackslash}m{0.2\linewidth} >{\centering\arraybackslash}m{0.2\linewidth}}
    \phantom{MMMMMMM} & \phantom{MMMMMMM} & \phantom{MMMMMMM} & \phantom{MMMMMMM} \\
    \multicolumn{4}{c}{\includegraphics[width=0.8\linewidth]{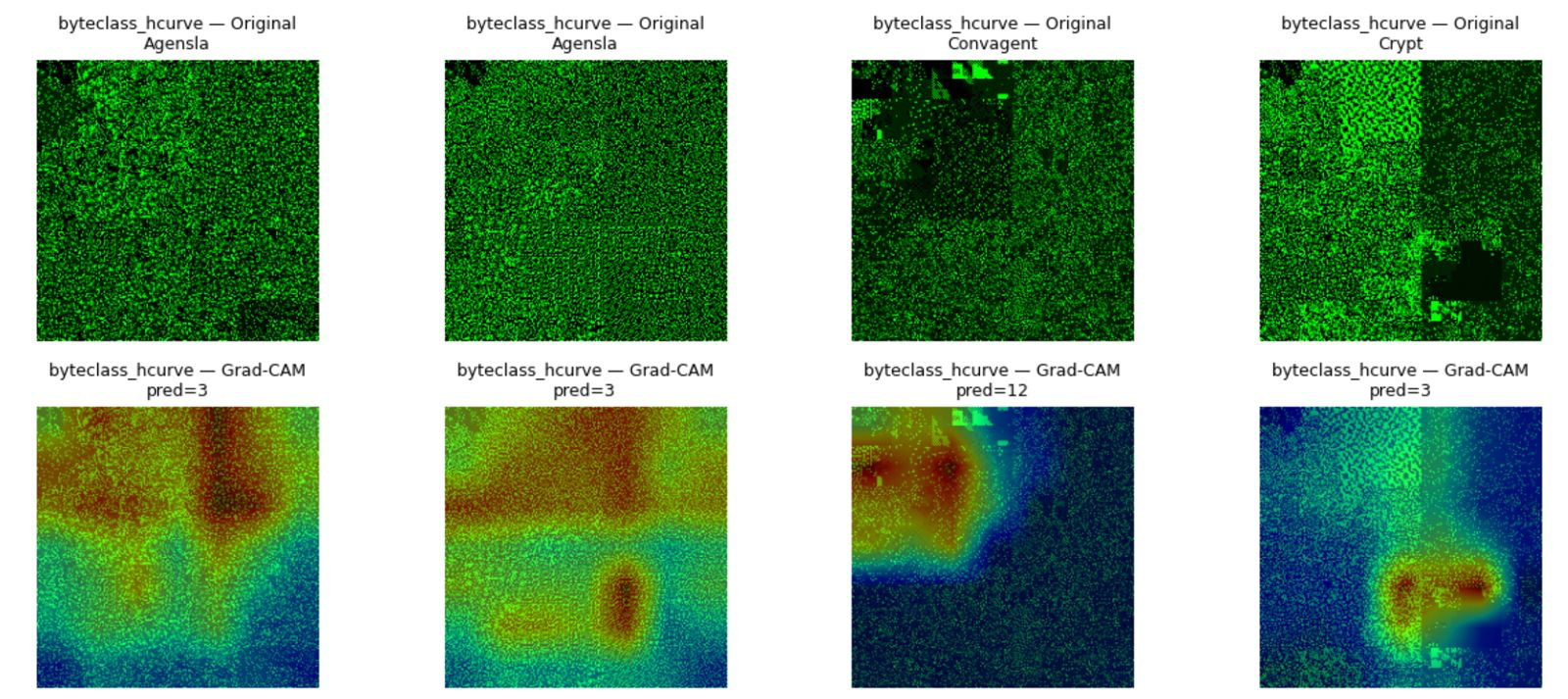}} \\[-1ex]
    \adjustbox{scale=0.85}{\ \ \ \ \texttt{Agensla}} & \adjustbox{scale=0.85}{\ \ \ \ \texttt{Agensla}} 
    	& \adjustbox{scale=0.85}{\ \ \texttt{Convagent}} & \adjustbox{scale=0.85}{\texttt{Crypt}} \\
%    \adjustbox{scale=1.0}{\ \ \ \ \ \ \ \ \texttt{Agensla}} & \adjustbox{scale=1.0}{\ \ \ \ \texttt{Agensla}} 
%    	& \adjustbox{scale=1.0}{\texttt{Convagent}} & \adjustbox{scale=1.0}{\!\!\!\!\!\!\!\!\texttt{Crypt}} \\
    \end{tabular}
    \caption{\byteclassHilbert\ malware images and Grad-CAM overlays}\label{fig:gradcam2b}
\end{figure}

\begin{figure}[!htb]
    \centering
    \begin{tabular}{>{\centering\arraybackslash}m{0.2\linewidth} >{\centering\arraybackslash}m{0.2\linewidth} 
    	>{\centering\arraybackslash}m{0.2\linewidth} >{\centering\arraybackslash}m{0.2\linewidth}}
    \phantom{MMMMMMM} & \phantom{MMMMMMM} & \phantom{MMMMMMM} & \phantom{MMMMMMM} \\
    \multicolumn{4}{c}{\includegraphics[width=0.8\linewidth]{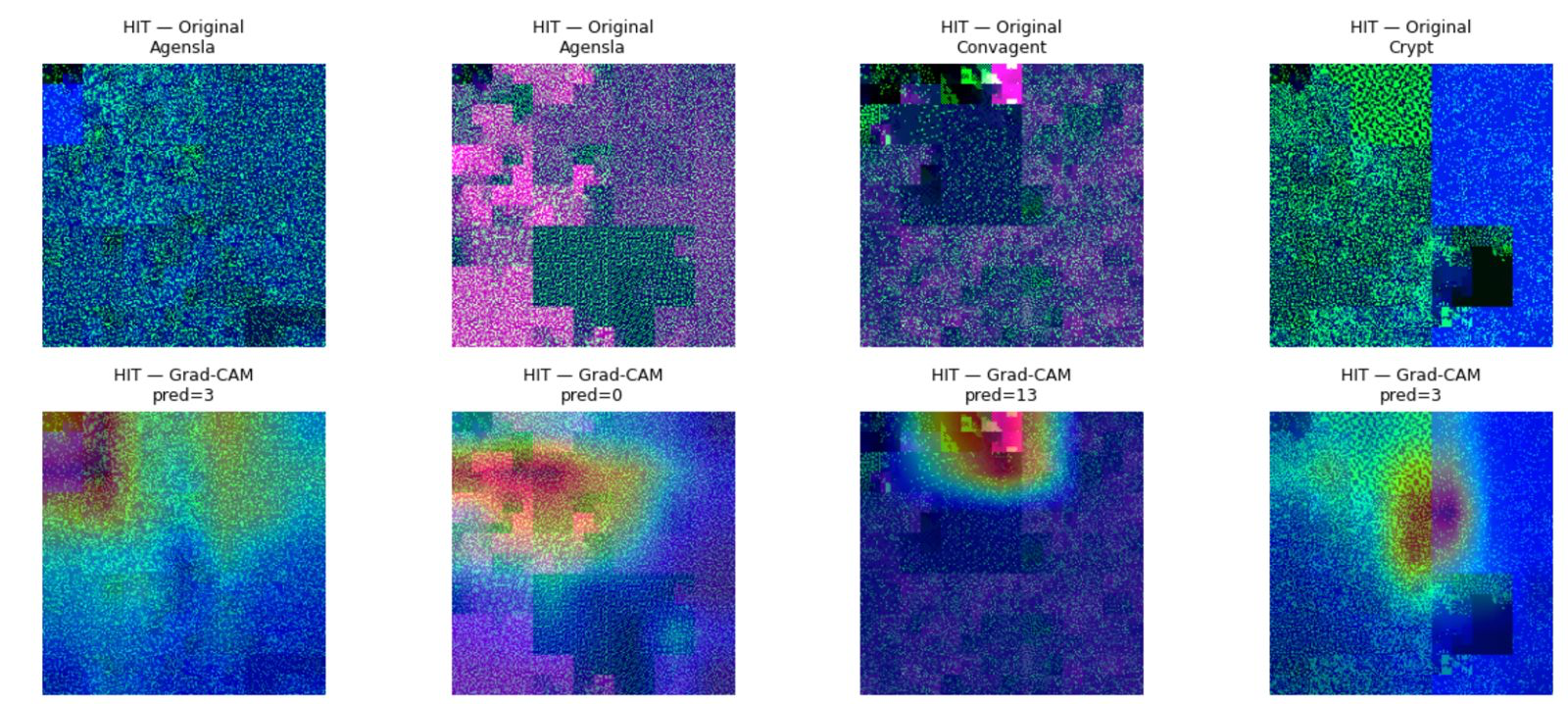}} \\[-1ex]
    \adjustbox{scale=0.85}{\ \ \ \ \texttt{Agensla}} & \adjustbox{scale=0.85}{\ \ \ \ \texttt{Agensla}} 
    	& \adjustbox{scale=0.85}{\ \ \texttt{Convagent}} & \adjustbox{scale=0.85}{\texttt{Crypt}} \\
%    \adjustbox{scale=1.0}{\ \ \ \ \ \ \ \ \texttt{Agensla}} & \adjustbox{scale=1.0}{\ \ \ \ \texttt{Agensla}} 
%    	& \adjustbox{scale=1.0}{\texttt{Convagent}} & \adjustbox{scale=1.0}{\!\!\!\!\!\!\!\!\texttt{Crypt}} \\
    \end{tabular}
    \caption{\hit\ malware images and Grad-CAM overlays}\label{fig:gradcam2c}
\end{figure}

\begin{figure}[!htb]
    \centering
    \begin{tabular}{>{\centering\arraybackslash}m{0.2\linewidth} >{\centering\arraybackslash}m{0.2\linewidth} 
    	>{\centering\arraybackslash}m{0.2\linewidth} >{\centering\arraybackslash}m{0.2\linewidth}}
    \phantom{MMMMMMM} & \phantom{MMMMMMM} & \phantom{MMMMMMM} & \phantom{MMMMMMM} \\
    \multicolumn{4}{c}{\includegraphics[width=0.8\linewidth]{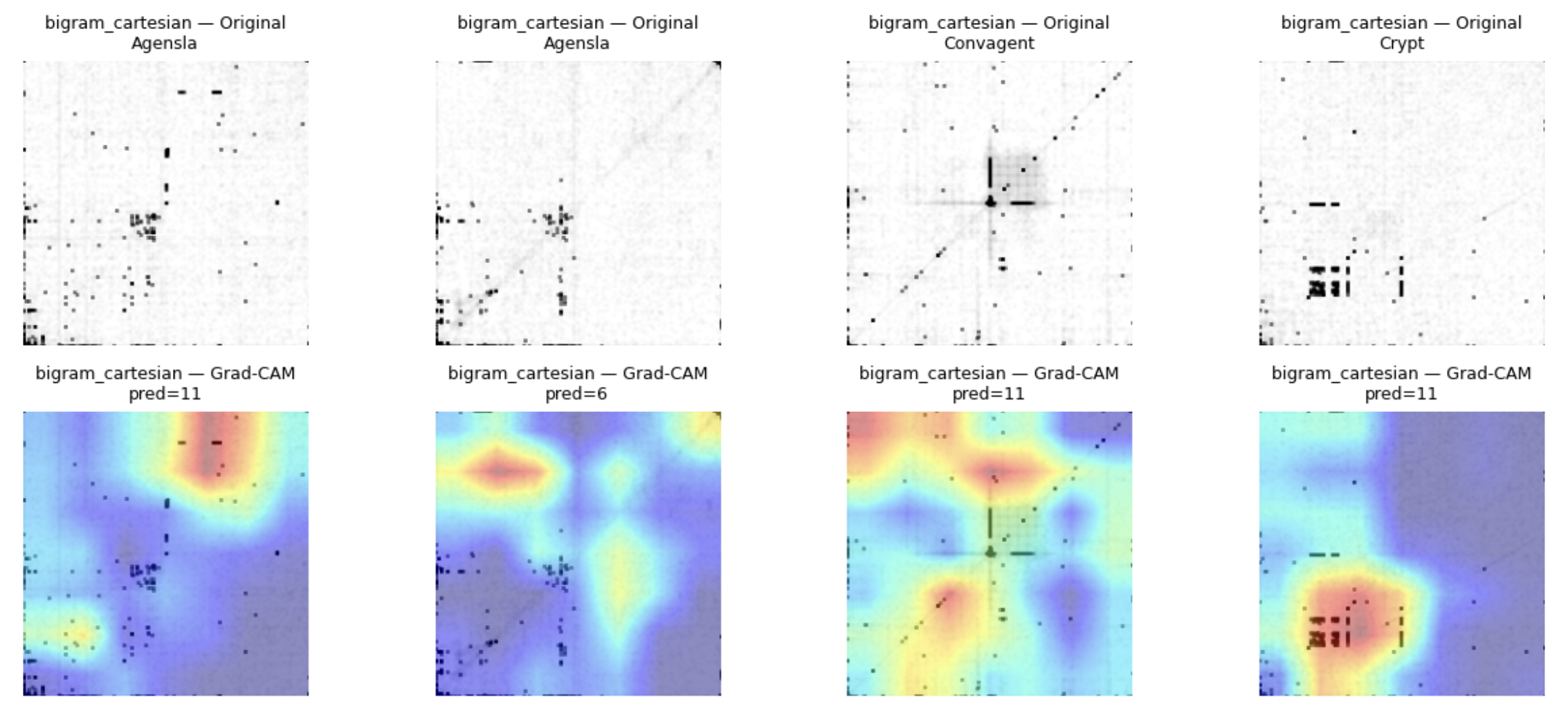}} \\[-1ex]
    \adjustbox{scale=0.85}{\ \ \ \ \texttt{Agensla}} & \adjustbox{scale=0.85}{\ \ \ \ \texttt{Agensla}} 
    	& \adjustbox{scale=0.85}{\ \ \texttt{Convagent}} & \adjustbox{scale=0.85}{\texttt{Crypt}} \\
%    \adjustbox{scale=1.0}{\ \ \ \ \ \ \ \ \texttt{Agensla}} & \adjustbox{scale=1.0}{\ \ \ \ \texttt{Agensla}} 
%    	& \adjustbox{scale=1.0}{\texttt{Convagent}} & \adjustbox{scale=1.0}{\!\!\!\!\!\!\!\!\texttt{Crypt}} \\
    \end{tabular}
    \caption{\bigramCartesian\ malware images and Grad-CAM overlays}\label{fig:gradcam2d}
\end{figure}

\begin{figure}[!htb]
    \centering
    \begin{tabular}{>{\centering\arraybackslash}m{0.2\linewidth} >{\centering\arraybackslash}m{0.2\linewidth} 
    	>{\centering\arraybackslash}m{0.2\linewidth} >{\centering\arraybackslash}m{0.2\linewidth}}
    \phantom{MMMMMMM} & \phantom{MMMMMMM} & \phantom{MMMMMMM} & \phantom{MMMMMMM} \\
    \multicolumn{4}{c}{\includegraphics[width=0.8\linewidth]{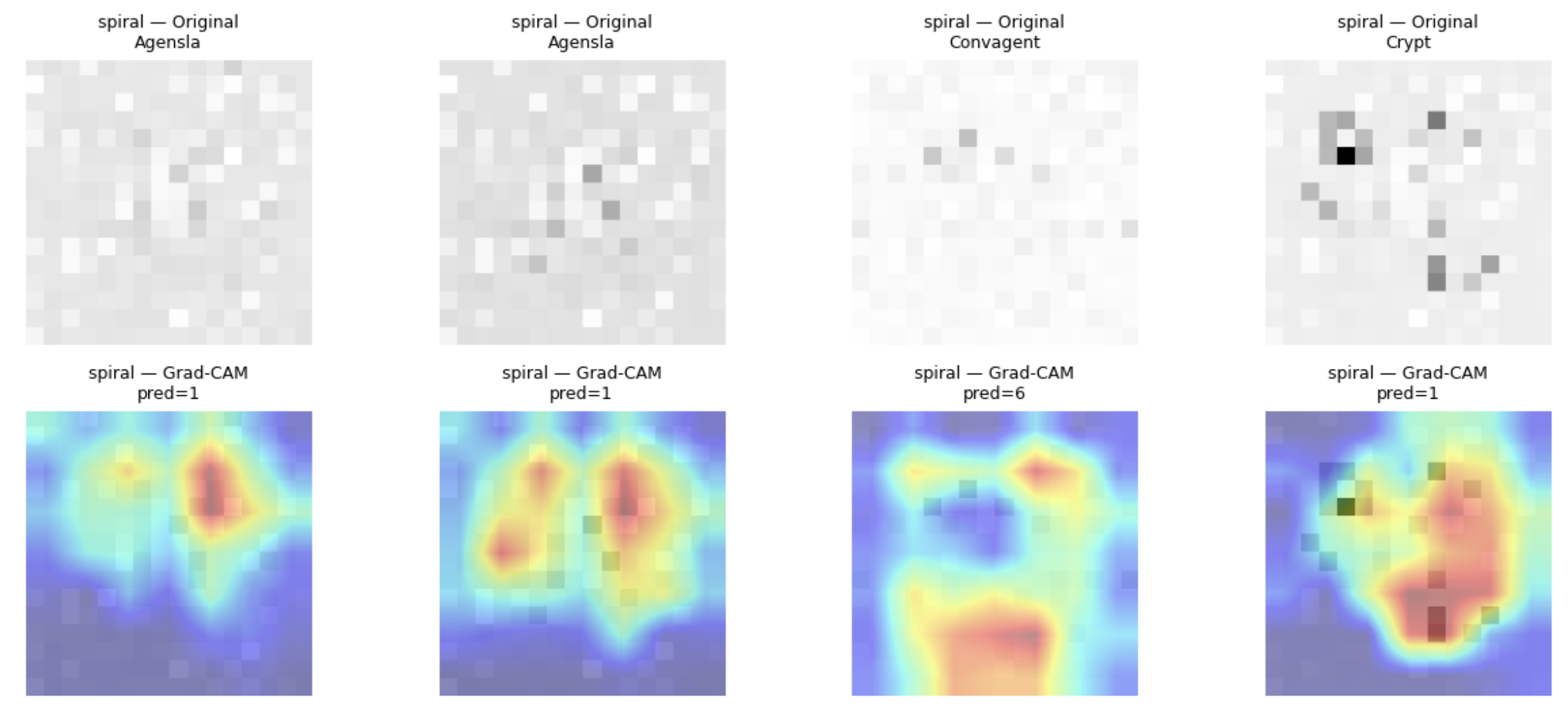}} \\[-1ex]
    \adjustbox{scale=0.85}{\ \ \ \ \texttt{Agensla}} & \adjustbox{scale=0.85}{\ \ \ \ \texttt{Agensla}} 
    	& \adjustbox{scale=0.85}{\ \ \texttt{Convagent}} & \adjustbox{scale=0.85}{\texttt{Crypt}} \\
%    \adjustbox{scale=1.0}{\ \ \ \ \ \ \ \ \texttt{Agensla}} & \adjustbox{scale=1.0}{\ \ \ \ \texttt{Agensla}} 
%    	& \adjustbox{scale=1.0}{\texttt{Convagent}} & \adjustbox{scale=1.0}{\!\!\!\!\!\!\!\!\texttt{Crypt}} \\
    \end{tabular}
    \caption{\spiral\ malware images and Grad-CAM overlays}\label{fig:gradcam2e}
\end{figure}

\begin{figure}[!htb]
    \centering
    \begin{tabular}{>{\centering\arraybackslash}m{0.2\linewidth} >{\centering\arraybackslash}m{0.2\linewidth} 
    	>{\centering\arraybackslash}m{0.2\linewidth} >{\centering\arraybackslash}m{0.2\linewidth}}
    \phantom{MMMMMMM} & \phantom{MMMMMMM} & \phantom{MMMMMMM} & \phantom{MMMMMMM} \\
    \multicolumn{4}{c}{\includegraphics[width=0.8\linewidth]{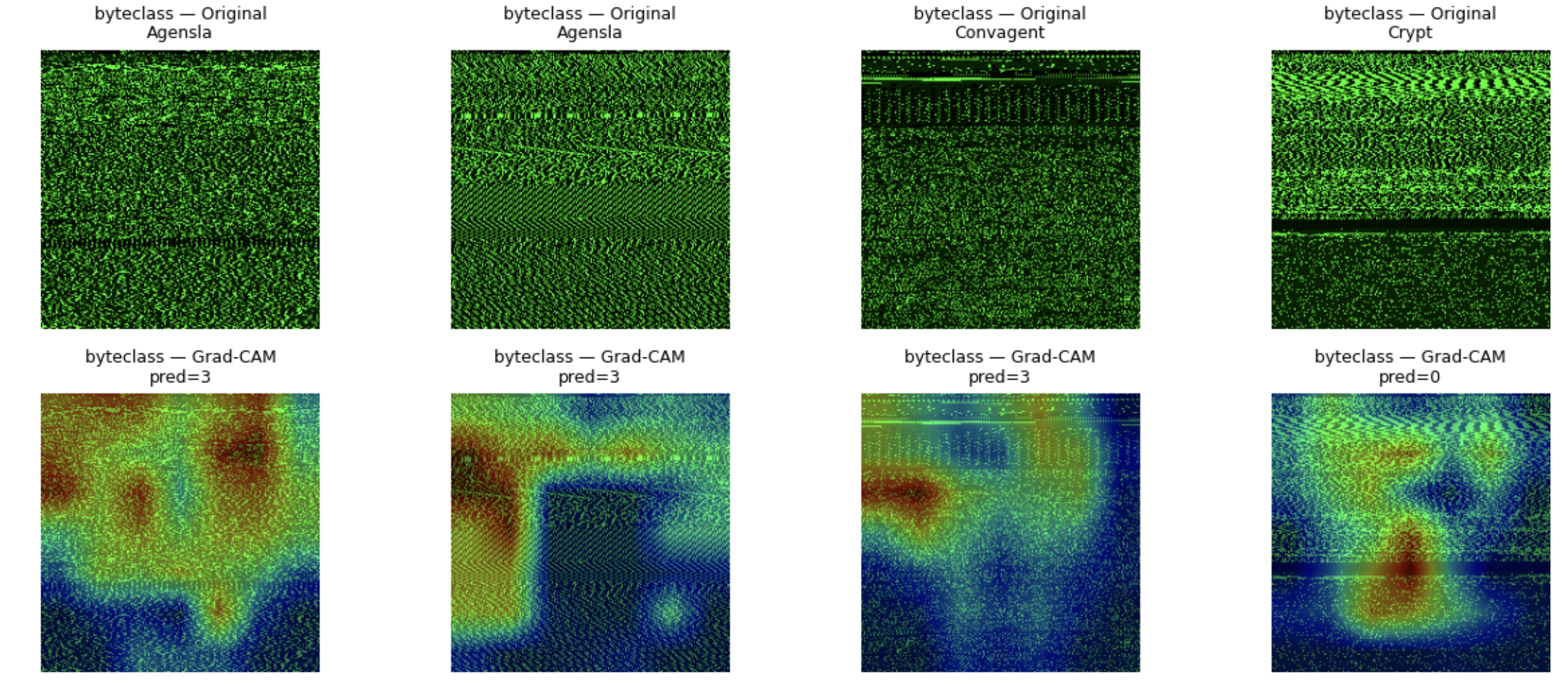}} \\[-1ex]
    \adjustbox{scale=0.85}{\ \ \ \ \texttt{Agensla}} & \adjustbox{scale=0.85}{\ \ \ \ \texttt{Agensla}} 
    	& \adjustbox{scale=0.85}{\ \ \texttt{Convagent}} & \adjustbox{scale=0.85}{\texttt{Crypt}} \\
%    \adjustbox{scale=1.0}{\ \ \ \ \ \ \ \ \texttt{Agensla}} & \adjustbox{scale=1.0}{\ \ \ \ \texttt{Agensla}} 
%    	& \adjustbox{scale=1.0}{\texttt{Convagent}} & \adjustbox{scale=1.0}{\!\!\!\!\!\!\!\!\texttt{Crypt}} \\
    \end{tabular}
    \caption{\byteclass\ malware images and Grad-CAM overlays}\label{fig:gradcam2f}
\end{figure}

In this Appendix, we give examples of malware images and their corresponding 
Grad-CAM overlay images. Of the eight image conversion types considered in
this chapter, \entropyHilbert, \byteclassHilbert, \hit, \bigramCartesian, \spiral, 
and \byteclass\ appear here in Figures~\ref{fig:gradcam2a} through~\ref{fig:gradcam2f}, respectively.
Analogous examples for the \grayscale\ and \bigramPolar\ image conversion
types are given, respectively, in Figures~\ref{fig:gradcam}(a) and~\ref{fig:gradcam}(b) 
in Section~\ref{sect:GC}.

%\clearpage

\end{document}